\documentclass[preprint,sort&compress,12pt]{elsarticle}
\usepackage{bbm}
\usepackage{amstext}
\usepackage{amsmath}
\usepackage{amssymb}
\usepackage{mathrsfs}
\usepackage{stmaryrd}
\usepackage{bm}
\usepackage{pstricks}
\usepackage{pst-coil}
\usepackage{pst-3d}
\usepackage{amsfonts}
\usepackage{amsthm}
\usepackage{CJK}

\newcommand{\mm}{\mathrm}
\newcommand{\ml}{\mathcal}

\newcommand{\dx}{\mathrm{d}x}

\newcommand{\be}{\begin{equation}}
\newcommand{\bea}{\begin{equation}\begin{aligned}}
\newcommand{\beas}{\begin{equation*}\begin{aligned}}
\newcommand{\eeas}{\end{aligned}\end{equation*}}
\newcommand{\eea}{\end{aligned}\end{equation}}
\newcommand{\ee}{\end{equation}}

\renewcommand{\div}{{\rm div }}

\begin{document}
\begin{CJK*}{GBK}{song}
\begin{frontmatter}
\title{
On Magnetic Inhibition Theory
in  Non-resistive Magnetohydrodynamic Fluids}

\author[FJ]{Fei Jiang}
\ead{jiangfei0591@163.com}
\author[sJ]{Song Jiang}
\ead{jiang@iapcm.ac.cn}
\address[FJ]{College of Mathematics and
Computer Science, Fuzhou University, Fuzhou, 350108, China.}
\address[sJ]{Institute of Applied Physics and Computational Mathematics, P.O. Box 8009,
 Beijing, 100088, China.}

\begin{abstract}
We investigate why the non-slip boundary condition for the velocity, imposed in the direction of impressed magnetic fields, can contribute to the magnetic
 inhibition effect based on the magnetic Rayleigh--Taylor (abbr. NMRT) problem in nonhomogeneous incompressible non-resistive magnetohydrodynamic (abbr. MHD) fluids.
 Exploiting an infinitesimal method in Lagrangian coordinates, the idea of (equivalent) magnetic tension, and the differential version of magnetic flux
 conservation, we give an explanation of physical mechanism for the magnetic inhibition phenomenon in a non-resistive MHD fluid. Moreover,
  we find that the magnetic energy in the non-resistive MHD fluid depends on the displacement of fluid particles, and thus can be regarded
  as elastic potential energy. Motivated by this observation, we further use the well-known minimum potential energy principle to
  explain the physical meaning of the stability/instability criteria in the NMRT problem. As a result of the analysis, we further extend
  the results on the NMRT problem to the stratified MHD fluid case. We point out that our magnetic inhibition theory can be used to explain
  the inhibition phenomenon of other flow instabilities, such as thermal instability, magnetic buoyancy instability, and so on, by impressed magnetic
  fields in non-resistive MHD fluids.
\end{abstract}
\end{frontmatter}


\newtheorem{thm}{Theorem}[section]
\newtheorem{lem}{Lemma}[section]
\newtheorem{pro}{Proposition}[section]
\newtheorem{concl}{Conclusion}[section]
\newtheorem{cor}{Corollary}[section]
\newproof{pf}{Proof}
\newdefinition{rem}{Remark}[section]
\newtheorem{definition}{Definition}[section]

\section{Introduction}\label{introud}
\numberwithin{equation}{section}

The study of the inhibition of flow instability by (impressed) magnetic fields goes back to the theoretical work of Chandrasekhar, who first discovered the inhibiting effect of a sufficiently large (impressed) vertical magnetic field on the thermal (or convective) instability
  based on the linearized magnetic Boussinesq equations of magnetohydrodynamic (abbr. MHD) fluids in a horizontal layer domain in 1952 \cite{CSTPRSOTICB,CSTPRSOTICBII}. Then Nakagawa experimentally verified Chandrasekhar's linear magnetic inhibition theory in 1955 \cite{YNAEN,YNAEN2}.
  Later, many authors tried to provide a rigorous mathematical proof of the magnetic inhibiting theory for the nonlinear case.
  In 1985, Gladi first successfully showed the theory for the nonlinear magnetic Boussinesq equations with resistivity
  by using a so-called generalized energy method \cite{GGPNA}. Now, it still remains open to mathematically
  show Chandrasekhar's assertion that thermal instability could be also inhibited
  by a vertical (magnetic) field in non-resistive MHD fluids, see \cite[page 160]{CSHHSCPO}. An alternative question arises whether
  one can mathematically verify the inhibition of other instabilities by a magnetic field in non-resistive  MHD fluids.
  The answer is positive, for example, the authors of this paper recently have mathematically verified the magnetic inhibition phenomenon
  (or stability result) in the nonhomogeneous magnetic Rayleigh--Taylor (abbr. NMRT) problem \cite{JFJSJMFMOSERT}.
  Such a result in the NMRT problem supports Chandrasekhar's assertion in the certain sense.

In \cite{JFJSJMFMOSERT}, the authors also showed the inhibition effect of a horizonal field for the case that the {NMRT} problem is considered
in a vertical layer domain. Moreover, the authors further pointed out that the nonslip boundary condition of the velocity at the two parallel fixed slabs,
imposed in the direction of the magnetic field, can contribute to the magnetic inhibition effect. This also gives a reason why the horizonal field does not
have inhibition effect as the vertical magnetic in the RT instability in a horizontal layer domain. Recently, the inhibition effect of a horizontal field is also found by the authors in the study of the magnetic buoyancy instability (the Parker instability) problem \cite{JFJSSETEFP}.

As mentioned above, progress has been made on the mathematical analysis of the magnetic inhibition phenomenon in non-resistive MHD fluids, however,
to our best knowledge, there is still no physical interpretation why the nonslip boundary condition, imposed in the direction of the (impressed)
magnetic field, can contribute to the magnetic inhibition effect. In this article, we investigate the physical mechanism of the magnetic inhibition theory
based on the {NMRT} problem. Next, we briefly introduce our main results.

First, we exploit an infinitesimal method in Lagrangian coordinates to give a compelling physical mechanism of the inhibition effect of magnetic
fields on the RT instability for the known mathematical results in the {NMRT} problem. The physical mechanism can be described as follows.
In the {NMRT} problem, we can think that the non-resistive MHD fluid under equilibrium is made up of infinite (fluid)
element lines which are parallel to the impressive field. Once we disturb the rest state, the element lines will be bend.
By using a differential version of magnetic flux conservation in Lagrange coordinates, we can compute out that the direction of the magnetic
field at each point of element lines is just tangent to the element lines in motion. Combining with the idea that the Lorentz force (a body force) can be equivalent
to a surface force, i.e., a so-called magnetic tension,
we can find that each element line can be regarded as an elastic string, and the magnetic tension intensity is proportional to the impressed field intensity.
Thus, the magnetic tension will resist gravity and straighten the all bent element lines, when the impressed field intensity is sufficiently large.
Noting that the endpoints of all element lines are fixed due to the non-slip boundary condition for the velocity,
thus all bent element lines will try to restore to their initial locations,
and can vibrate around their initial location under the magnetic tension. In particular, due to the viscosity, all bent element lines will
asymptotically converge to their initial location.
The corresponding details will be presented in Section \ref{201706140828}.
Moreover, we obtain a so-called equivalence theorem of magnetic flux conservation in the analysis process, see Theorem \ref{201706231335}.
We mention that our magnetic inhibition mechanism with nonslip boundary condition of the velocity can be also used
to explain the inhibition phenomenon of thermal instability and magnetic buoyancy instability  \cite{SKMRFNA} by magnetic fields.

Second, we further give the physical meaning of stability and instability criteria in the {NMRT} problem.
By the physical mechanism of the magnetic inhibition effect and the mathematical representation of magnetic energy, we
find that the magnetic energy in non-resistive MHD fluids can be regarded as the elastic potential energy.
This means that the well-known minimum potential energy principle can be applied to the {NMRT} problem.
More precisely, if the total potential energy in the magnetic energy and the gravitational potential energy in the rest state is minimal,
then the {NMRT} problem is stable. Otherwise, the {NMRT} problem is unstable. Motivated by the minimum potential energy principle, and using some mathematical techniques,
we can indeed prove that under the stability criterion, the total potential energy in the rest state reaches its minimum,
while under the instability criterion, the total potential energy in the rest state is not minimal, see Theorem \ref{201707191444}.
We shall also extend the results on the {NMRT} problem to the stratified magnetic RT (abbr. {SMRT}) problem (see Theorem \ref{201707191444n}), and
the corresponding analysis details will be presented in Section \ref{201706191309} and the rigorous proof in Section \ref{2017111222010}.

Finally, in Section \ref{Sec:0302}, we extend the mathematical result of the magnetic inhibition in the NMRT problem to the magnetic
Boussinesq problem without heat conduction, and shall see that the obtained result supports Chandrasekhar's assertion in the absence of heat conduction.

We end this section by listing some notations which will be used throughout this article.
\vspace{2mm}

(1) Basic notations:

 The superscript $\mm{T}$ denotes the transposition. $I$ always denotes the $3\times 3$ identity matrix, $e_i$ stands for the unit vector,
in which the $i$-th component is $1$. $\det A$ denotes the determinant of the matrix $A$. For $x:=(x_1,x_2,x_3)\in \mathbb{R}^3$,
 we define $x_{\mm{h}}:=(x_1,x_2)$ and $x_{\mm{v}}:=(x_2,x_3)$. $\mathcal{T}:=\mathbb{R}/\mathbb{Z}$ is the usual $1$-torus.
Let $f:=(f_1,f_2,f_3)^{\mm{T}}$ be a vector function defined in a three-dimensional domain, we define $f_{\mm{h}}:=(f_1,f_2)^{\mm{T}}$,
$\nabla_{\mm{h}}f_{\mm{h}}:=(\partial_j f_i)_{2\times 2}$, $\partial_{\vec{n}}f:=\vec{n}\cdot \nabla f$, and
  $\partial_{\vec{n}}^2 f:= (\vec{n}\cdot \nabla)^2 f$, where $\vec{n}$ is a constant vector.
$\mm{d}y_{\mm{h}}:=\mm{d}y_1\mm{d}y_2$ is the infinitesimal on a plane. $\varphi^0$ denotes the initial data of a scalar,
or a vector, or a matrix function $\varphi(x,t)$, where $t$ is the time variable.

(2) Definitions of domains and boundaries:
\begin{align}
& \mathbb{T}^2=(2\pi L_1\mathcal{T})\times( 2\pi L_2\mathcal{T}),\quad     \Omega^+:=\{x:=(x_1, x_{\mm{v}})\in \mathbb{R}^3~|~x_{\mm{v}}\in \mathbb{T}^2, 0<x_1<h\},\nonumber \\
&\nonumber \Omega_{a}^b:=\{x:=(x_{\mm{h}}, x_3)\in \mathbb{R}^3~|~x_{\mm{h}}\in \mathbb{T}^2, a< x_3<b\},\\ &\Omega_+:= \Omega_{0}^h,\ \Omega_-:= \Omega_{-l}^0,\quad  \Omega:=\Omega_+\cup  \Omega_-,\quad \Omega_a^b\!\!\!\!\!\!\!-\ :=\mathbb{R}^2\times (a,b),\nonumber\\
& \Sigma_{a}:=\mathbb{T}^2\times \{x_3=a\},\quad  \Sigma_+:=\Sigma_h ,\quad \Sigma:= \Sigma_{0},\quad \Sigma_-:= \Sigma_{-l},\nonumber
 \end{align}
where $L_1$, $L_2$, $h$, $l$, $a$, $b>0$, and $a>b$. Let $D$ be a domain, then $\overline{D}$ denotes the closure of $D$, and
$$\partial D\mbox{ denotes }\left\{
                              \begin{array}{ll}
                            \mbox{the boundary of }\Omega, & \hbox{ if }\Omega\mbox{ is a bounded domain}; \\
 \{x_1=0\}\cup \{x_1=h\} , & \hbox{ if }D=\Omega^+; \\
                        \{x_3=a\}\cup \{x_3=b\}, & \hbox{ if }D=\Omega_a^b.
                              \end{array}
                            \right.
$$
We define $f(P):=\{x\in \mathbb{R}^3~|~x=f(y)\mbox{ for }y\in P\}$, where the vector function $f$ is defined on the set $P$.
It is should be noted that if a function is defined on $\Omega_+$, then the function is horizontally periodic, i.e.,
$$f(x_1,x_2,x_3)=f(2m\pi L_1+x_1, 2n\pi L_2+x_2,x_3)\mbox{ for any integer }m\mbox{ and }n.$$
Similarly, if a function is defined on $\Omega^+$, then the function is vertically periodic function.

(3) Notations of function spaces and simplified norms:
\begin{align}
&{H}^i(D):=W^{i,2}(D)\mbox{ denotes a Soblev space defined in } D,\nonumber \\
& H^1_\sigma(D):=\{\eta\in {H}^1(D)~|~\div \eta=0,\ \eta|_{\partial D}=0\mbox{ in the sense of trace}\},
\nonumber\\
& C^\infty_\sigma(\Omega_a^b):=\{w\in {C}^\infty(\Omega_a^b)~|~w=0\mbox{ on }\Omega_a^b \setminus\Omega_c^d\mbox{ for some }\Omega_c^d\nonumber\\
& \qquad \qquad\   \quad \mbox{ satisfying }a<c<d<b,\ \div  w=0 \},\nonumber \\
& {H}_{0,*}^{k,1}(\Omega_+):=\{\varpi \in H^1_0(\Omega_+)\cap H^k(\Omega_+)~|~\varpi(y)+y:\overline{\Omega_0^h\!\!\!\!\!\!\!-\ } \mapsto \overline{\Omega_0^h\!\!\!\!\!\!\!-\ }\nonumber
\\ & \qquad\qquad\qquad \mbox{ is a homeomorphism mapping},\ \det(\nabla(\varpi (y)+y))=1\},\nonumber\\ & {H}_{0,*}^{k,1}(\Omega):=\{\varpi \in H^1_0(\Omega_{-l}^h)\cap H^k(\Omega)~|~\varpi(y)+y:\overline{\Omega_{-l}^h\!\!\!\!\!\!\!\!\!-\ }\ \mapsto \overline{\Omega_{-l}^h\!\!\!\!\!\!\!\!\!-\ }\nonumber
\\ & \qquad\qquad\qquad \mbox{ is a homeomorphism mapping},\ \det(\nabla(\varpi (y)+y))=1\},\nonumber\\ & \|\cdot \|_{i,D}:=\|\cdot \|_{H^i(D)},\quad |\cdot|_{i} := \|\cdot\|_{H^{i}(\mathbb{T}^2)},\nonumber
\end{align}
where $i\geqslant 0$, and $k\geqslant 2$ are positive constant. It should be noted that $W^{0,p}(D):=L^p(D)$ is the usual Lebesgue space.
In addition, if a norm is defined in a periodic domain or a horizontally/vertically periodic domain, then the norm is equal to the one
defined in a periodic cell. For examples, $\|\cdot \|_{H^k(\Omega_a^b)}=\|\cdot \|_{H^k((0,2\pi L_1)\times( 0,2\pi L_2)\times (a,b))}$,
and $\|\cdot \|_{H^k(\mathbb{T}^2)}=\|\cdot \|_{H^k((0,2\pi L_1)\times( 0,2\pi L_2) )}$.  Similarly, we have
$\int_{\Omega_a^b}=\int_{(0,2\pi L_1)\times( 0,2\pi L_2)\times (a,b)}$ and $\int_{\mathbb{T}^2}=\int_{(0,2\pi L_1)\times( 0,2\pi L_2)}$.

\section{Magnetic inhibition  mechanism}\label{201706140828}

This section is devoted to providing the  physical  mechanism of magnetic inhibition phenomenon. We first recall the known mathematical results
on the {NMRT} problem in Subsection \ref{201707142031} and define the direction of a (impressed) magnetic field in a non-resistive MHD fluid. We then
reformulate the momentum equations of the perturbed MHD fluid in Lagrangian coordinates in Subsection \ref{sec:02}.
Finally, we use the differential version of magnetic flux conservation to define the direction of the magnetic field,
and give thus the reason why the no-slip boundary condition of the velocity, imposed in the direction of the impressed field, can contribute
to the magnetic inhibition effect in Subsection \ref{201708291624}.

\subsection{Stability and instability results of the {NMRT} problem}\label{201707142031}

Let us first recall the stability and instability results of the {NMRT} problem. To start with,
we introduce the  three-dimensional homogeneous incompressible viscous MHD equations  with zero resistivity in the presence
of a uniform gravitational field in a domain $D\subset {\mathbb R}^3$,
\begin{equation}\label{comequations2151x}
\left\{\begin{array}{l}
\rho_t+v\cdot \nabla \rho=0,\\[1mm]
\rho v_t+\rho v\cdot\nabla v+\nabla
   p -\mu\Delta v=\lambda  (\nabla \times M)\times  M -\rho g e_3,\\[1mm]
M_t=M\cdot \nabla v-v\cdot\nabla M,\\[1mm]
\mathrm{div}v=\mathrm{div} {M}=0.\end{array}\right.\end{equation}
Here the unknowns $\rho:= \rho(x,t)$, ${v}:= {v}(x,t)$, $M:= {M}(x,t)$ and $p:= p(x,t)$ denote the density, velocity,
magnetic field and the kinetic pressure of a MHD fluid, respectively; the positive constants $\lambda$, $\mu$ and $g$ stand for
the permeability of vacuum divided by $4\pi$, the coefficient of shear viscosity and the gravitational constant,
respectively. For the system \eqref{comequations2151x}, we impose the initial and boundary conditions:
\begin{align}
&\label{0104}
(\rho ,v,N)|_{t=0}=(\rho^0,v^0,N^0)\ \mbox{in } D,\\[1mm]
&\label{0105nn} v(x,t)|_{\partial D}={  0}\ \mbox{ for any }t>0.
\end{align}
We call \eqref{comequations2151x}--\eqref{0105nn} the NMHD model.
We mention that the well-posedeness problem of viscous MHD equations with zero resistivity have been extensively investigated, see \cite{TZWYJGw,XXRZYXIAngZFZhang,RXXWJHXZYZZF,RHPYZYZHU,HXPGETDCMFLin,FHLinPzhang,FHLXLPZ10221526,XLPZ10221525} for examples.

Now we consider a uniform  magnetic field  (i.e., every component of the magnetic field is constant, and at least one component is non-zero)  $\bar{M}=(\bar{M}_1,\bar{M}_2,\bar{M}_3)$ and a smooth RT density profile $\bar{\rho}\in C^2(\overline{D})$,
which is independent of $(x_1,x_2)$ and satisfies
\begin{eqnarray}\label{0102}
\inf_{ x\in D}\bar{\rho}>0,\quad\;\bar{\rho}'|_{x_3=x^0_{3}}>0\quad\mbox{ for some }x^0_{3}\in {D_{x_3}}:= \{x_3~|~(x_{\mm{h}},x_3)\in D\}, \end{eqnarray}
where we have denoted $\bar{\rho}':=\mm{d}\bar{\rho}/\mm{d}x_3$.
 Then, ${r_{\mm{N}}}:=(\bar{\rho},0,\bar{M})$ is a rest state (or equilibrium) solution of the NMHD model with an associated pressure
  $\bar{p}$ defined by the following relation
\begin{equation}
\label{201706151213}
\nabla \bar{p}=-\bar{\rho}g  {e}_3.
\end{equation}
We often call $\bar{M}$ the impressed magnetic field.
The problem whether the rest state ${r_{\mm{N}}}$ is stable or unstable to the NMHD model is called the {NMRT} problem.

The second condition in \eqref{0102} is called the RT condition and assures that there is at least a region in which the RT density profile
has larger density with increasing height $x_3$, thus leading to the RT instability under small perturbation for a sufficiently small $ {\bar{M}}_3$.
However, for a sufficiently large $|\bar{M}|$, the RT instability may be inhibited. Recently, the authors have shown the inhibition
of the RT instability by a magnetic field \cite{JFJSJMFMOSERT}. Before recalling the results in \cite{JFJSJMFMOSERT}, we introduce some notations.

Denote $\Pi_1:=(\bar{M}_3,0,0)^{\mm{T}}$, $\Pi_3:=(0,0,\bar{M}_3)^{\mm{T}}$, $\Omega_1=\Omega^+$ and $\Omega_3=\Omega_+$. If $D$ takes $\Omega^+$,
we shall further assume that the density profile $\bar{\rho}$ is a vertically horizontal function, i.e.,
$$\bar{\rho}|_{x_3=2\pi n L_2}=\bar{\rho}|_{x_3=2\pi m L_2}\;\;\mbox{ for any integer }n,m.$$
We define
\begin{equation*}
\mathscr{V}_{g\bar{\rho}'}^D(w_3):= g\int_{D}\bar{\rho}'w_3^2\dx\quad\mbox{and}\quad
\mathscr{V}_{\vec{n}}^{D}(w):=\lambda\|\partial_{\vec{n}} w\|^2_{0,D},
\end{equation*}
where $\vec{n}$ is a constant vector. We further define
\begin{align}
\label{201711041321}
   m_{\mm{N},j}^{D}:=\sqrt{\sup_{w\in H_{\sigma}^1(D)}\frac{\mathscr{V}_{g\bar{\rho}'}^D(w)}{\mathscr{V}_{e_j}^{D}(w)}}.
\end{align}
We remark here that if $j=3$ and $D=\Omega_+$, we rewrite $m_{\mm{N},3}^{\Omega_+}$ by $m_{\mm{N}}$ for the sake of simplicity.

Then the authors have established the following stability and instability results for the {NMRT} problem \cite{JFJSJMFMOSERT}:
\begin{enumerate}[\quad (1)]
\item Stability criterion: if
\begin{equation*}
|\bar{M}_j|>{m}_{\mm{N},j}^{\Omega_j}\;\;\mbox{ (called asymptotic stability condition)},
\end{equation*}
then the rest state  $r_{\mm{N}}$ with $\bar{M}=\Pi_j$ is asymptotically stable to the {NMRT} model defined on $D=\Omega_j$
with proper initial conditions under small perturbation for $j=1$ and $3$.
\item Instability criterion:  if
\begin{equation*}
|\bar{M}_j|<{m}_{\mm{N},j}^{\Omega_j}\;\;\mbox{ (called instability condition),}
\end{equation*}
then the rest state  ${r_{\mm{N}}} $  with $\bar{M}=\Pi_j$  is unstable to the {NMRT} model defined on $D=\Omega_j$ in the Hadamard sense for $j=1$ and $3$.
In addition, the rest state ${r_{\mm{N}}}$ with $\bar{M}=\Pi_k$ is always unstable to the {NMRT} model defined on $D=\Omega_j$ in the Hadamard sense, if $k\neq j$.
\end{enumerate}

Since $\bar{\rho}$ satisfies the RT condition and the velocity is non-slip on the boundary of $\Omega_j$, we can verify that
$m_{\mm{N},j}^{\Omega_j}\in (0,\infty)$. Thus the above positive constant ${m}_{\mm{N},j}^{\Omega_j}$ is called a strength-threshold of $\Pi_j$
for stability and instability of the {NMRT} problem. Moreover, for $\bar{\rho}'$ being a positive constant, we can compute out that
\begin{equation}
\label{20170902}
{m}_{\mm{N},j}^{\Omega_j}=2 h\sqrt{g \bar{\rho}'/\sigma\pi}.\end{equation}

Considering the case $D=\Omega_+$, we find that the rest state ${r_{\mm{N}}}$ with $\bar{M}=(\bar{M}_1,0,0)^{\mm{T}}$ is always unstable to
the {NMRT} model defined on $\Omega_3$. This means that a horizontal field can not inhibit the RT instability in a horizontally periodic domain.
However, in view of the stability criterion for $j=1$, we find that a horizontal magnetic field can inhibit the RT instability in a vertically
periodic domain with finite width. Thus, one observes that the non-slip boundary condition, imposed in the direction of the magnetic field,
can contribute to the magnetic inhibition effect.

Of course, we can use the threshold \eqref{201711041321} to explain the above fact.
Considering the case $D=\Omega_+$ and $\bar{M}=(0,0,\bar{M}_3)$,
for any $w\in H^1_\sigma(\Omega_+)$, since $w|_{\partial\Omega_+}=0$ and $\Omega_+$ is horizontally periodic with finite height,
then one has $\| w_3\|_{0,\Omega_+}^2\leqslant c   \|\partial_3w_3\|_{0,\Omega_+}^2$ for some constant $c$ depending on the domain, which implies
 $m_{\mm{N} } \in (0,\infty)$. However, for any positive constat $j$, we can always construct a function $w\in H_\sigma^1(\Omega_+)$ satisfying
 $\| w_3\|_{0,\Omega_+}^2>j   \|\partial_1 w_3\|_{0,\Omega_+}^2$, which
 implies $m_{\mm{N},1}^{\Omega_+}=\infty$. Thus, the strength of any horizontal field $\bar{M}$ is always less than $m_{\mm{N},1}^{\Omega_+}$, and
 this shows why the rest state ${r_{\mm{N}}}$ with $\bar{M}=  (\bar{M}_1,0,0)$ is always
 unstable to the {NMRT} model defined $\Omega_+$.

Unfortunately, from the above mentioned stability/instability results we can not see any physical mechanism to explain
why the non-slip boundary condition can contribute to the magnetic inhibition effect.
To reveal the physical mechanism of the magnetic inhibition phenomenon,
we shall carry out analysis of forces based the equations of non-resistive MHD fluids in Lagrangian coordinates.

\subsection{Reformulation in Lagrangian coordinates}\label{sec:02}

From now on, we always assume that $\bar{M}$ is a general non-zero uniform magnetic field. However, to avoid a discussion of the geometry structure
of a general domain $D$ in the analysis of (fluid) element lines, we only consider the simplest domain, i.e., $D=\Omega_+$. Of course,
the magnetic inhibition mechanism for $D=\Omega_+$ can be easily generalized to a general domain that is bounded in the direction of $\bar{M}$.

We assume that $(\rho,v,M,q)$ be a classical solution to the NMHD model defined on $Q_+^T:=\Omega_+\times [0,T]$ with $T>0$, in which $v$ satisfies
the following regularity
\begin{equation}\label{2017711102319}
 v\in C^1(\overline{Q_+^T})\;\;\mbox{ and }\;\;\nabla^2 v\in C^0(\overline{Q_+^T}).
\end{equation}

It is well-known that the (generalized) Lorentz force on the right-hand side of \eqref{comequations2151x}$_2$
can be written as the divergence of the magnetic part of the electromagnetic stress, i.e.,
\begin{equation}
\label{201706091719}
\lambda(\nabla \times M)\times  M =\lambda\mm{div}(  M \otimes M-|M|^2I/2) .
\end{equation}
We consider a bounded domain $V$ with smooth surface in $\Omega_+$, then the Lorentz force acting on the
 MHD fluid in $V$ is given by the formula
\begin{equation}
\label{2017060917191451}\lambda \int_V (\nabla \times M)\times  M  \mm{d}x= \lambda \int_S F_\nu\mm{d}S,
\end{equation}
where \begin{equation*}
F_\nu=\lambda M\cdot\nu M-\lambda |M|^2\nu/2
\end{equation*} and $\nu$ denotes the unit outer normal vector of $V$.  The
first term $\lambda M\cdot\nu M$ in $F_\nu$ is called the (equivalent) magnetic tension, the direction of which is along that of the magnetic field,
and the strength of which is $\lambda|M|^2$. The second term $-\lambda |M|^2\nu/2 $ is called the isotropic (equivalent) magnetic pressure. The relation \eqref{2017060917191451} indicates that the Lorentz force (a body force) can be equivalent to a surface force  \cite{BJAFP315200}.

Now we use the relation \eqref{201706091719} to rewrite \eqref{comequations2151x}$_2$  as follows:
\begin{equation}
\label{201706091542}
\rho v_t+\rho v\cdot\nabla v+\nabla
p^* -\mu\Delta v =\lambda \mm{div}( M \otimes M)-\rho g e_3  ,
\end{equation}
where $p^*$ denotes the sum of the kinetic pressure $p$ and the magnetic pressure $\lambda  |M|^2/2$. Hence, one can think that the momentum equation
  \eqref{201706091542} describes the motion of a fluid driven by the magnetic tension $\lambda M\cdot\nu M$ and the gravity force $-\rho g e_3$.
We denote the equations \eqref{comequations2151x} with \eqref{201706091542} in place of \eqref{comequations2151x}$_2$ by the system
\eqref{comequations2151x}$^*$.

In view of the system \eqref{comequations2151x}$^*$,
it is very important to analyze how the magnetic tension affects the motion of the MHD fluid.  To this end, we
shall rewrite \eqref{comequations2151x}$^*$ in Lagrange coordinates.
We first label the particles (or element points) of the non-resistive fluid by the relation $y=x$, where $x$ denote the Eulerian coordinates
of particles at $t=0$, and we call $y$  the Lagrange coordinates (or Lagrangian particle markers).
Then we define a location function (or particle-trajectory mapping) $\zeta$  of the fluid particles $y$ as the solution to
\begin{equation*}
\left\{
            \begin{array}{l}
 \zeta_t (y,t)=v(\zeta(y,t),t)
\\
\zeta (y,0)=y,
                  \end{array}    \right.
\end{equation*}
where $y\in \overline{\Omega_+}$. Obviously, $\eta:=\zeta-y$ represents the displacement function of particles.
By the regularity \eqref{2017711102319} and the classical ODE theory \cite{WWODE148}, the solution $\zeta$ enjoys the following regularity:
\begin{equation*}
\zeta\in C^1(\overline{Q_+^T})\mbox{ and } \zeta \in C^2(Q_+^T).
\end{equation*}
Note that the non-slip boundary condition \eqref{0105nn} with $\Omega_+$ in place of $D$ is essential here,
since it guarantees that for each $t\in [0, T]$, $\zeta(\cdot,t):\Omega_+\to \Omega_+$ is a homeomorphism mapping (referring to Lemma \ref{lem:modfied}).

Before giving the motion equations \eqref{comequations2151x} in Lagrangian coordinates, we temporarily introduce some notations
involving $\zeta$.  Define $\mathcal{A}:=(\ml{A}_{ij})_{3\times 3}$ via
\begin{equation}
\label{201712091617}
\ml{A}^{\mm{T}}=(\nabla \zeta)^{-1}:=(\partial_j \zeta_i)^{-1}_{3\times 3},
\end{equation}
$J:=\det(\nabla \zeta)$, and the differential operators $\nabla_{\ml{A}}$ and $\mm{div}_\ml{A}$ as follows.
\begin{align}
&\nabla_{\ml{A}}w:=(\nabla_{\ml{A}}w_1,\nabla_{\ml{A}}w_2,\nabla_{\ml{A}}w_3)^{\mm{T}},\quad \nabla_{\ml{A}}w_i:=(\ml{A}_{1k}\partial_kw_i,
\ml{A}_{2k}\partial_kw_i,\ml{A}_{3k}\partial_kw_i)^{\mm{T}},\nonumber \\
&\mm{div}_{\ml{A}}(f_1,f_2,f_3)^{\mm{T}}=(\mm{div}_{\ml{A}}f_1,\mm{div}_{\ml{A}}f_2,\mm{div}_{\ml{A}}f_3)^{\mm{T}},
\quad \mm{div}_{\ml{A}}f_i:=\ml{A}_{lk}\partial_k f_{il},\nonumber \\
& \Delta_{\mathcal{A}}w:= (\Delta_{\mathcal{A}}w_1,\Delta_{\mathcal{A}}w_2,\Delta_{\mathcal{A}}w_3)^{\mm{T}}\;\;\mbox{ and }\;\;
\Delta_{\mathcal{A}}w_i:=\mm{div}_{\ml{A}}\nabla_{\ml{A}}w_i\nonumber
\end{align}
for vector functions $w:=(w_1,w_2,w_3)^{\mm{T}}$ and $f_i:=(f_{i1},f_{i2},f_{i3})^{\mm{T}}$,
where we have used the Einstein convention of summation over repeated indices, and $\partial_{k}$ denotes the partial derivative
with respect to the $k$-th component of $y$. Moreover, $\zeta$ enjoys the following properties:
\begin{enumerate}[\qquad (1)]
\item Let $\delta_{ij}$ be the Kronecker delta, then
\begin{equation}
\label{201707161455}
\partial_i\zeta_k\ml{A}_{kj}= \ml{A}_{ik}\partial_k\zeta_j=\delta_{ij}.
\end{equation}
\item Since $\mm{div}v=0$, one has $J=J^0$, where $J^0$ is the initial value of $J$. In particular, if $J^0=1$, then
\begin{equation}\label{zetanabla}J =1 .
\end{equation}
\item In view of the definition of $\mathcal{A}$ and \eqref{zetanabla}, we can see that $\mathcal{A}=(A^*_{ij})_{3\times 3}$, where $A^{*}_{ij}$ is the algebraic complement minor of the $(i,j)$-th entry $\partial_j \zeta_i$. Moreover, we can compute out that $ \partial_kA^{*}_{ik}=0$, which implies
\begin{equation}\label{diverelation}
\mm{div}_{\mathcal{A}}u=\partial_l ( {A}_{kl}^*u_k)=0.
\end{equation}
\end{enumerate}

Now, we define the Lagrangian unknowns by
\begin{equation*}
(\varrho, u,q^*,B)(y,t)=(\rho,v,p^*,M)(\zeta(y,t),t)\;\;\mbox{ for } (y,t)\in \Omega_+\times\mathbb{R}^+.
\end{equation*}
Thus in Lagrangian coordinates, the evolution equations for ${\varrho}$, $u$, $B$ and $q^*$ read as
\begin{equation}\label{0116}
\left\{
\begin{array}{ll}
\zeta_t=u, \\[1mm]
{\varrho}_t=0,\\
{\varrho} u_t+\nabla_{\ml{A}}q^*-\mu \Delta_{\ml{A}}u=
\lambda B\cdot \nabla_{\ml{A}} B- {\varrho} ge_3, \\[1mm]
B_t-B\cdot\nabla_{\ml{A}}u=0, \\[1mm]
\div_\ml{A}u=\div_\ml{A}B=0.
\end{array}
\right.
\end{equation}
Since we slightly disturb the rest state only in the velocity at the initial time $t=0$, we have
\begin{equation}
\label{201707161156}
(\zeta,{\varrho}, u, B)|_{t=0}=(y,\bar{\rho}, v^0, \bar{M}).
\end{equation}
We mention that $\mm{div}_{\mathcal{A}}B=0$ automatically holds since the initial value of $\mm{div}_{\mathcal{A}}B $ is just zero.
In addition, in view of the non-slip boundary condition of $v$, we obtain the following boundary condition:
\begin{equation}
\label{201712100806}
(\zeta,u)|_{\partial\Omega_+}=(y,0).
\end{equation}

\subsection{Differential version of magnetic flux conservation}\label{201708291624}

The most important advantage of using Lagrange coordinates lies in that the magnetic field can be expressed
by the location function $\zeta$, so that we can determine the direction of the magnetic tension. Next, we derive this fact.

Consider the equations
\begin{equation}\label{0116n} \left\{
                              \begin{array}{ll}
\zeta_t=u, \\[1mm]
 {\varrho}_t=-\varrho\mm{div}_{\mathcal{A}}u,  \\[1mm]
B_t-B\cdot\nabla_{\ml{A}}u=-B \mm{ div}_{\mathcal{A}}u
\end{array}
                            \right.
\end{equation}
with initial condition $(\zeta,\varrho,B)|_{t=0}=(\zeta^0,\varrho^0,B^0)$. We should note here that
 $\mm{div} u$ may not be $0$ in \eqref{0116n}. It is well-known that the system \eqref{0116n} can
imply the mass conservation $J\varrho=J^0\varrho^0$ and magnetic flux conservation
\begin{eqnarray}
\label{0124sdfs}
J\mathcal{A}^{\mm{T}} B = J^0\ml{A}^{\mm{T}}_0 B^0,
\end{eqnarray}
which can be rewritten as a so-called Cauchy's integral of the magnetic field \cite{SDPTSPSCR}:
\begin{eqnarray}
\label{0124}   B=J^0\nabla\zeta \ml{A}^{\mm{T}}_0 B^0/J.
\end{eqnarray}
Here we call \eqref{0124sdfs} the magnetic flux conservation, since we can directly verify that the conservation relation \eqref{0124sdfs}
is equivalent to the well-known theorem of magnetic flux conservation in non-resistive MHD fluids, see Theorem \ref{201706231335}
in the next subsection.

For an incompressible non-resistive MHD fluid, since $\div_\ml{A}u=0$, one obtains $J=J^0$ in Lagrangian coordinates.
Hence, \eqref{0124} reduces to
\begin{eqnarray}
\label{0124nnn}   B= \nabla\zeta \ml{A}^T_0 B^0.
\end{eqnarray}
The above formula can be also directly deduced from the equations \eqref{0116}$_1$ and \eqref{0116}$_4$, and
we also call \eqref{0124nnn} the vorticity-transport  formula in the theory of the vorticity equation, see \cite[Propotition 1.8]{MAJBAL}.
Since $B|_{t=0}=\bar{M}$ and $\mathcal{A}_0^{\mm{T}}=I$, we further deduce from \eqref{0124nnn} that
\begin{equation}
\label{201706151104}
 B=\partial_{\bar{M}} \zeta.
\end{equation}
The above expression has the prominent physical meaning that the larger the stretching $\nabla \zeta$ of a non-resistive MHD fluid along the direction
of $\bar{M}$ is, the stronger the magnetic field becomes. Next, we use \eqref{201706151104} to determine the direction of the magnetic tension
of each particle in motion.

Let us think that the fluid is made up of infinite element lines, which are parallel to $\bar{M}$. We slightly disturb the rest state
in the velocity by a perturbation $v^0$ at $t=0$. The motion equations in Lagrangian coordinates after perturbation are described by \eqref{0116} with
initial-boundary value conditions \eqref{201707161156} and \eqref{201712100806}. Now, consider a straight element line denoted by $l$ at $t=0$.
We denote by the set $l^0$ (it is a segment for $M_3\neq 0$, and a line without endpoints for $M_3=0$) the initial location occupied by $l$.
Under a slight perturbation, the element line $l$ may be bent, and may move to a new location at time $s$, which is occupied by $l$ and
denoted by the set $l^s$ (it is a curve, if $l$ is bent). We further choose a particle $Y$ on $l$ at $t=0$, and denote the coordinates
of $Y$ at $t=0$ and $t=s$ by $y^0$ and $x^s$, respectively. Hence, $x^s=\zeta(y^0,s)$.

Noting that the curve $l^s$ can be defined by the location function $\zeta(y,s)$ defined on $l^0$, one can easily verify
that $\partial_{\bar{M}}\zeta(y^s,s)$ is a tangential direction at point $x^s$ of the curve $l^s$. In other words, the curve $l^s$ is tangent
to the direction of the magnetic field at each point. Recalling the definition of the magnetic line (i.e., a line which is tangent to the direction
of the magnetic field at each point, and the direction of a magnetic line can be defined by that of the magnetic field),
we find that $l^s$ is just a magnetic line. This is just the well-known (magnetic field)
line conservation theorem (or loosely called the frozen-in magnetic field lines in some literature \cite{LZXLY}),
i.e., any two particles in a non-resistive MHD fluid that are at an instant on a common magnetic field line will keep
 on the common magnetic field line at any other instant \cite{SDPTSPSCR,FCGCO}.

Since $l^s$ is a magnetic field line, by the mathematical expression of the magnetic tension, we see that the element line $l$ at any time $s$
can be always regarded as an elastic string due to the magnetic tension along the curve $l^s$.  Thus, if the element line is bent at time $s$,
then the magnetic tension will drive the bend part of the element line back to a straight line,
thus playing a role of restoring force and representing the stabilizing effect in the motion of the non-resistive MHD fluid.
We take the following figure to illustrate this fact by infinitesimal method.

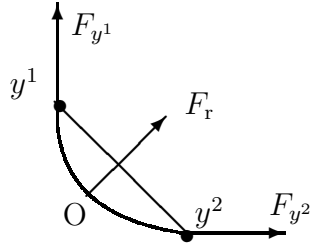
\begin{center}
\begin{picture}(160,95)(0,0)
\thicklines
\put(48,50){$\bullet$}
\put(32,58){$y^1$}
\put(50,52){\vector(0,1){40}}
\put(55,80){$F_{y^1}$}
\put(96,1){$\bullet$}
\put(102,9){$y^2$}
\put(96,5){\vector(1,0){40}}
\put(130,12){$F_{y^2}$}
\put(52,8){O}
\put(61,19){\vector(1,1){30}}
\put(98,50){$F_{\mm{r}}$}
\qbezier(50,55)(47,12)(98,5)
\put(49,55){\line(1,-1){50}}
\end{picture}
\label{2017110011040}
\end{center}

The element curve $y^1Oy^2$ in the above figure is a bend part, denoted by $l_{\mm{p}}$, of element line $l$ at time $s$, and the  element segment $y^1y^2$
is the initial location of $l_{\mm{p}}$. We assume that the direction of $\bar{M}$ is parallel to the vector $\overrightarrow{y^2y^1}$.
Now, we denote by $F_{Y^i}$ the magnetic tension acting on the element endpoint $Y^i$ of $l_{\mm{p}}$ at time $s$ for $i=1$ and $2$. Noting that
$y^1Oy^2$ is a magnetic field line, in view of the expression of the magnetic tension, we see that the total magnetic tension acting on the element curve
$y^1Oy^2$ without two element endpoints $Y^1$ and $Y^2$ is zero, and the magnetic tension acting on the element endpoint $Y^i$ is given by the formula
$$F_{y^i}=\lambda  \partial_{\bar{M}} \zeta(y^i,s)\cdot \nu_{y^i}   \partial_{\bar{M}} \zeta(y^i,s),$$
where $\nu_{y^i}$ denotes the unit outer normal vector at element endpoint $y^i$ for $i=1$ and $2$.
Consequently, the magnetic tension acting on the bend element curve $y^1Oy^2$ is just the resultant force $F_{\mm{r}}$ of $F_{y^1}$ and $F_{y^2}$ from the above figure. Obviously, the resultant force $F_{\mm{r}}$ drives the element curve $y^1Oy^2$ to recover to a straight element line. From the above figure and \eqref{201706151104},
we can intuitively see that the more bent the magnetic line $y^1Oy^2$ is, the stronger the resultant force $F_{\mm{r}}$ (or called
the magnetic restoring force) becomes.

By the above analysis on the physical mechanism of stabilizing effect of magnetic fields, we can easily explain the stability/instability phenomenon
in the {NMRT} problem under small perturbation. First, we consider the case $\bar{M}_3\neq 0$. Since the fluid is incompressible and $v$ is zero
on the boundary, there always exist some bent element lines after disturbing the rest state. Once some element lines are bent, the destabilizing
factor (i.e. gravity) may promote the development of the RT instability, which will further make the element lines more bent.
Though the magnetic tension resists the bend of the element lines, it can not prevent the heavier fluid from sinking under gravity,
if it is too small. Hence, we shall increase the strength of $\bar{M}$ so that a properly large magnetic tension can resist gravity\footnote{In Theorem \ref{201707191444} in the next section, we shall see that only the increment $|\bar{M}_3|$ of $\bar{M}$ contributes
to the inhibition effect. Here we provide an explanation for this. Assume that $y^1$ and $y^2$ in the above figure
are fixed on the upper and lower boundaries of $\Omega_+$, respectively. For the slightly bent case in the above figure, one can
intuitively see that the stronger the magnetic restoring force acting on the element curve $y^1Oy^2$ is, the bigger the intensity of $\bar{M}$ is.
However, we can also easily observe that the bigger the value $|\bar{M}_i|$ ($i=1$ or $2$) is, the longer the length of $y^1Oy^2$ is.
This means that one can not improve the intensity of the magnetic restoring force acting on the unit length of $y^1Oy^2$
by increasing the value of $\bar{M}_i$. This explains why the threshold is independent of $|M_i|$ in $\Omega_+$.}.
Noting that the endpoints of all element lines are fixed, one sees that all bent element lines
will recover to their initial location, and may vibrate around their initial location under the magnetic tension with sufficiently large $\bar{M}_3$\label{201711031343},
and the magnetic tension represents the inhibition effect. Consequently, under the effect of viscosity, all bent elements will asymptotically converge to their initial location.

Now, we roughly explain why the small height $h$ of the domain also contributes to stability, see \eqref{20170902}. To clearly see the reason,
we consider a particle $Y^{\mm{m}}$ which just lies at the midpoint on the element line $l$ at $t=0$. The fixed two endpoints of $l$ are labeled
by $y^{i}$ for $i=1$ and $2$. We assume that there exists a force $F$ pulling $Y^{\mm{m}}$ along the perpendicular bisector of $l^0$. So,
  $Y^{\mm{m}}$ moves to a new location $y^s$ at time $s$. Obviously, the magnetic line $y^1y^sy^2$ is more bent if $h$ is smaller.
  As aforementioned, the more bent the magnetic line is, the stronger the magnetic restoring force becomes.
Thus, if $h$ is getting smaller, the magnetic line $y^1y^sy^2$ will result in a stronger magnetic restoring force, which will resist
the pulling force $F$. This means that when the height $h$ is getting smaller, it is more difficult for the RT instability in a non-resistive MHD fluid to occur.
In addition, for the two-dimensional case, by the magnetic inhibition mechanism we can easily guess that a sufficiently large $(\bar{M}_1,0)$ also has magnetic inhibition effect
in the domain $\mathbb{T}\times (0,h)$.
More precisely, if one disturbs the rest state, i.e., $(\rho,M, v)=(\bar{\rho}(x_2),\bar{M},0)$ in the velocity,
then the density, magnetic field and velocity will finally converge to $(\bar{\rho}(x_2),\bar{M},0)$, but the location function may converge to some
non-zero function $(\zeta_1(x_2),x_2)$, where $\zeta_1(x_2)$ only depends on $x_2$, and $\zeta_1(x_2)|_{x_2=0,h}=0$.

Now we consider the case $\bar{M}_3=0$. Since the domain is horizontally periodic, points on the element lines can move freely.
 Thus, there may exist some element lines that are far away from their initial location and parallel to $\bar{M}$.
 Since such element lines are parallel to $\bar{M}$, the magnetic tension can not pull the element lines back to their initial location.
 Therefore, a horizonal magnetic field has no inhibition effect upon the RT instability in the horizontally periodic domain $\Omega_+$.
 Of course, if $\Omega_+$ is a vertically horizontal domain with finite width, then a horizonal magnetic field also has inhibition effect.

From the above analysis on both $\bar{M}_3\neq0$ and $\bar{M}_3=0$ cases, we have seen why the non-slip boundary condition for the velocity,
imposed in the direction of magnetic fields, can contribute to the inhibition effect. Moreover, we can intuitively conclude that the magnetic
inhibition should exist in a general bounded domain, if one component of $\bar{M}$ is sufficiently large, although the authors have no idea
to prove such a conclusion mathematically.

In the analysis of magnetic inhibition for $\bar{M}=(0,0,\bar{M}_3)^{\mm{T}}$, we consider the non-slip boundary condition ${u}|_{\partial\Omega_+}=0$.
An interesting question is that what will happen for the case that particles on $\partial\Omega_+$ can freely slip on the boundary of $\partial\Omega_+$,
i.e., the condition $u_3|_{\partial\Omega_+}=0$ is kept only. We guess that, if the element lines are slightly bent and $\bar{M}_3$ is sufficiently large,
then, due to the magnetic tension and viscosity, it could be possible that the bent element lines finally become straight again. Unfortunately,
we can not verify this mathematically.

\subsection{Equivalence theorem in magnetic flux conservation}

We further discuss the conservation relation \eqref{0124sdfs}. The conservation relation is not only very useful in the investigation of
the dynamic behavior of a magnetic field \cite{SDPTSPSCR}, but also in the study of some mathematical problems from MHD fluids, such as
the global well-posdness problem \cite{TZWYJGw,HXPGETDCMF,ABIHZPOTG}, and the inhibition effect duo to magnetic fields upon flow instabilities,
see \cite{WYC,JFJSWWWOA,JFJSNS,JFJSSETEFP,JFJSJMFMOSERT,WYTIVNMI}. As aforementioned, we can consider the conservation relation
as a differential version of magnetic flux conservation due to their equivalence. Next, we establish the equivalence theorem of magnetic flux
conservation in Eulerian coordinates and \eqref{0124sdfs}.

We begin with recalling the magnetic flux conservation in Eulerian coordinates. We choose a smooth surface $S^0$ in a MHD fluid in the rest state.
When the MHD fluid flows, the particles on the surface $S^0\subset \Omega_+ $ will form a new surface denoted by $S(t)$ at time $t$.
The equation of $S(t)$ can be given by
\begin{equation}
\label{201712092159}
S(t)=\{x\in \mathbb{R}^3~|~x=\zeta(y,t)\mbox{ for }y\in S^0~\}.
\end{equation}
In particular, $S(0)=S^0$. Then the magnetic flux passing through the surface $S(t)$ at $t$ can be formally given by the following formula:
\begin{equation}
\label{201712092111}
\int_{S(t)} M(x,t)\cdot\nu(x,t) \mm{d}S,
\end{equation}
and the magnetic flux conservation in Eulerian coordinates formally reads as follows   \cite{BJAFP315200}.
\begin{equation*}
 \int_{S(t)} M(x,t)\cdot\nu(x,t) \mm{d}S=\int_{S^0} \bar{M}\cdot\nu(x,0) \mm{d}S,
 \end{equation*}
where $\nu$ denotes the unit normal vector. However, each point on $S(t)$ has two unit normal vectors,
this results in a non-unique representation of the magnetic flux defined by \eqref{201712092111}.
Therefore, we shall use the theory of surfaces to refine the definition \eqref{201712092111}, so that $\nu$ can be definitely computed out.
To this end, we next introduce the definition of a regular surface.

A surface $S^0$ in $\Omega $ is called a regular surface, if  $S^0$ enjoys the following two properties:
\begin{itemize}
  \item  The surface $S^0$ can be parameterized by $y=f(\alpha,\beta)$ for $(\alpha,\beta)\in D^{S^0}$,
        where $D^{S^0}$ is a bounded and closed domain.
  \item  The function $y=f^{S^0}(\alpha,\beta)$: $D^{S^0}\to S^0$ is one to one, and belongs to $C^1(D^{S^0})$.
         Moreover, the corresponding Jacobi matrix, i.e.,
  $$\displaystyle \mathcal{J}\left(\frac{\partial f^{S^0}}{\partial \alpha}, \frac{\partial f^{S^0}}{\partial \beta}\right)=\left(
                  \begin{array}{ccc}
   \displaystyle  \frac{\partial f_1^{S^0}}{\partial \alpha},
        &   \displaystyle        \frac{\partial f_2^{S^0}}{\partial \alpha}, & \displaystyle          \frac{\partial f_3^{S^0}}{\partial \alpha} \\ [1em]
    \displaystyle   \frac{\partial f_1^{S^0}}{\partial \beta} &     \displaystyle \frac{\partial f_2^{S^0}}{\partial \beta} &    \displaystyle  \frac{\partial f_3^{S^0}}{\partial \beta}  \\
                  \end{array}
                \right)^{\mm{T}},
  $$ is full rank at every point of $ D^{S^0}$.
\end{itemize}

Now we define $\omega:=\zeta(f^{S^0}(\alpha,\beta),t)$. Then the Jacobi matrix of $\omega$ can be given by
$$\mathcal{J}\left(\frac{\partial \omega}{\partial \alpha}, \frac{\partial \omega}{\partial \beta}\right)=\nabla \zeta\mathcal{J}\left(\frac{\partial f^{S^0}}{\partial \alpha}, \frac{\partial f^{S^0}}{\partial \beta}\right).$$
Keeping in mind that $\zeta$ satisfies
\begin{align}
\label{201712092128}
&\zeta(y,0)=y\mbox{ in }\Omega_+,\quad \zeta(\cdot,t):\Omega_+\mapsto \Omega_+\;\mbox{ is one to one and belongs to }\; C^1(\Omega_+),\\
\label{201712092337}
&\det(\nabla \zeta(\cdot, t))\neq 0\mbox{ on }\Omega_+ \;\mbox{ for any given }\; t\in [0,T) ,
\end{align}
we immediately see that, for any given $t$, $S(t)$ is also a regular surface with the parameterized equation
$f^{S(t)}:=\zeta(f^{S^0}(\alpha,\beta),t)$ defined on $D^{S^0}$, and $\nu$ defined by the following formula is a normal vector
of each point $x$ on $S(t)$:
\begin{equation}
\label{201712092308}
\nu(x,t):= \partial_\alpha\zeta(f^{S^0}(\alpha,\beta),t)\times \partial_\beta\zeta(f^{S^0}(\alpha,\beta),t))/| \partial_\alpha\zeta(f^{S^0}(\alpha,\beta),t)\times \partial_\beta\zeta(f^{S^0}(\alpha,\beta),t))|
\end{equation}
for $t\in [0,T)$, where $\alpha$ and $\beta$ satisfies $x=f^{S^0}(\alpha,\beta)$.

From now on, we stipulate that once we choose a parameterized equation $f^{S^0}$ of the surface $S^0$,
then the normal vector of each point $x$ on $S(t)$ is defined by \eqref{201712092308}.
Then, we can show the following general equivalence theorem.
\begin{thm}
\label{201706231335}
Let $D$ be a domain, $\zeta$ satisfy \eqref{201712092128}--\eqref{201712092337}
with $D$ in place of $\Omega_+$, and let $J=\det{\ml{A}}$ with $\ml{A}$ being defined by \eqref{201712091617}. Assume
that $M(\cdot,t)\in C^0(D)$ for any $t\in (0,T)$, $M(\cdot,0)=\bar{M}$,
$B:=M(\zeta,t)$ and the initial value $\zeta(y,0)=y$.
\begin{itemize}[\quad (1)]
  \item For any given sub-domain $D_1\subseteq D$, if
  \begin{equation} \label{abjlj0newn}
J\ml{A}^{\mm{T}}B= \bar{M}\; \mbox{ in } \; D_1 \;\; \mbox{ for some }t\in (0,T)
 \end{equation}
 then, for any regular surface $S^0\subset D_1$ with a parameterized equation $f^{S^0}(\alpha,\beta)$ defined on $D^{S^0}$,  we have
\begin{equation}
\label{201706101101new}
 \int_{S(t)} M(x,t)\cdot\nu(x,t) \mm{d}S= \int_{S^0} \bar{M}\cdot \nu(x,0) \mm{d}S,
\end{equation}
where $S(t)$ is defined by \eqref{201712092159}, and $\nu(x,t)$ is defined by \eqref{201712092308}.
  \item[\quad (2)] Let $D_1\subseteq D$, if, for any bounded plane domain $S^0\subset D_1$, it holds that
 \begin{equation}
\label{201706101101newdafds}
 \int_{S(t)} M(x,t)\cdot\nu(x,t)\mm{d}{S}= \int_{S^0} \bar{M}\cdot\nu(x,0) \mm{d}S \quad\mbox{for some }t\in (0,T),
\end{equation}
where $S(t)$ is defined by \eqref{201712092159}, then $J\ml{A}^{\mm{T}}B= \bar{M}$ in $D_1$ at time $t$.
\end{itemize}
\begin{rem} Theorem \ref{201706231335} also holds for the case that $D$ is a closed domain, or a horizontally periodic domain. \end{rem}
\end{thm}
\begin{pf}
(1) We prove the first assertion. Since $\nu$ is defined by \eqref{201712092308},
 exploiting the integral formula on surfaces, we know that
\begin{align}
&\int_{S(t)} M(x,t)\cdot\nu(x,t) \mm{d}S\nonumber \\
&= \int_{D^{S^0}} M(\zeta(f^{S^0}(\alpha,\beta),t),t)\cdot(\partial_\alpha\zeta(f^{S^0}(\alpha,\beta),t)\times \partial_\beta\zeta(f^{S^0}(\alpha,\beta),t)) \mm{d} {S}.\label{201706171738}
\end{align}

Due to
$$
\begin{aligned}
&\partial_\alpha\zeta(f^{S^0}(\alpha,\beta),t)\times \partial_\beta\zeta(f^{S^0}(\alpha,\beta),t)\\
&= (\partial_2\zeta\times \partial_3\zeta,\partial_3\zeta\times \partial_1\zeta,
\partial_1\zeta\times \partial_2\zeta)|_{y=f^{S^0} (\alpha,\beta)}(\partial_\alpha f^{S^0}\times \partial_\beta f^{S^0})\\
&= (J\mathcal{A})|_{y=f^{S^0}(\alpha,\beta)}(\partial_\alpha f^{S^0}\times \partial_\beta f^{S^0}),
\end{aligned}
$$
we insert the above relation into \eqref{201706171738} to deduce
\begin{align}
\int_{S(t)} M(x,t)\cdot\nu(x,t) \mm{d}S= &\int_{D^{S^0}} (J M(\zeta(y,t),t))|_{y=f^{S^0}(\alpha,\beta)}\cdot(\mathcal{A}|_{y=f^{S^0}(\alpha,\beta)}(\partial_\alpha f^{S^0}\times \partial_\beta f^{S^0})) \mm{d} {S}\nonumber \\
=& \int_{D^{S^0}} (J\mathcal{A}^{\mm{T}}M(\zeta(y,t),t))|_{y=f^{S^0}(\alpha,\beta)}\cdot(\partial_\alpha f^{S^0}\times \partial_\beta f^{S^0}) \mm{d} {S}.
\label{201706171738nn}
\end{align}
Noting that $S^0\subset D_1$, we make use of \eqref{abjlj0newn} to get
$$ (J\mathcal{A}^{\mm{T}}M(\zeta(y,t),t))_{y=f^{S^0}(\alpha,\beta)}=\bar{M}\mbox{ in }D^{S^0}.$$
Recalling $\zeta(y,0)=y$, one has $\nu(x,0)=\partial_\alpha f^{S^0}\times \partial_\beta f^{S^0}/|\partial_\alpha f^{S^0}\times \partial_\beta f^{S^0}|$.
Thus, one further finds that
$$ \int_{S(t)} M(x,t)\cdot\nu(x,t) \mm{d}S= \int_{D^{S^0}} \bar{M} \cdot (\partial_\alpha f^{S^0}\times \partial_\beta f^{S^0})  \mm{d}  {S}=  \int_{S^0}\bar{M}\cdot \nu(x,0) \mm{d} \vec{S}.$$
Hence, \eqref{201706101101new} holds.

(2) Now we verify the second assertion.
Let $S^0\subset D_1$ be a bounded plan domain that is perpendicular to $x_3$-axis. Then there is a constant $a$, such that
$S^0\subset \{x\in \mathbb{R}^3~|~x_3=a\}$. Then we can choose a parameterized equation $f^{S^0}$ of $S^0$ as follows.
$$  f^{S^0}:=f^{S^0}(\alpha,\beta) =  (\alpha,\beta,a), $$
where $(\alpha,\beta)\in S^0_{x_3=a}:=\{(x_1,x_2)~|~(x_1,x_2,a)\in S^0 \}$.
Noting that $\partial_\alpha f^{S^0}\times \partial_\beta f^{S^0}=e_3$,
we can utilize \eqref{201706101101newdafds} and \eqref{201706171738nn} to further have
 $$  \begin{aligned}
&\int_{S^0} J\mathcal{A}^{\mm{T}}B  \cdot e_3 \mm{d} {S}\\
&=\int_{D^{S^0}}( J\mathcal{A}^{\mm{T}}M(\zeta(y,t),t))_{y=f^{S^0}(\alpha,\beta)} \cdot e_3 \mm{d} {S}=\int_{S^0} \bar{M} \cdot e_3\mm{d}S.
\end{aligned}$$
 In view of arbitrariness of $S^0\subset D_1$, we obtain immediately
$$(J\ml{A}^{\mm{T}}B)e_3= \bar{M}e_3\;\;\mbox{ in }\;D_1\;\mbox{ at time }\; t.$$

Similarly, we can also show that for $i=1$ and $2$,
$$(J\ml{A}^{\mm{T}}B)e_i= \bar{M}e_i\;\; \mbox{ in }\; D_1\; \mbox{ at time }\; t.$$
 Consequently, we have $J\ml{A}^{\mm{T}}B= \bar{M}$ in $D_1$ at time $t$. This completes the proof of the desired conclusion.
\hfill$\Box$
\end{pf}

\section{Physical interpretation of the stability and instability criteria}\label{201706191309}

In this section, we give a physical interpretation for the stability and instability criteria in the {NMRT} problem,
and then extend the obtained results for the {NMRT} problem to the {SMRT} problem.
We recall here again that $\bar{M}$ is a general uniform magnetic field in this section.

\subsection{Case of the {NMRT} problem}\label{201711021003}

We proceed to discuss the physical interpretation for the stability and instability criteria in the NMRT problem.
We start with the discussion of the physical meaning of $\mathscr{V}_{\bar{M}}^{\Omega_+}(\eta(t))$ and $\mathscr{V}_{g\bar{\rho}'}^{\Omega^+}(\eta(t))$.

It is well-known that the total magnetic energy $\mathscr{E}_M^{\Omega_+}(t)$ (defined on a periodic cell) of a MHD fluid in domain
$\Omega_+$ at the time $t$ is given by the formula:
$$\mathscr{E}_M^{\Omega_+}(t) =\frac{\lambda}{2}\int_{\Omega_+} |M|^2(x,t)\mm{d}x.$$
Making use the transform of Lagrange coordinates, \eqref{zetanabla} and the magnetic field expression \eqref{201706151104}, we see that $$\mathscr{E}_M^{\Omega_+}(t)=\frac{\lambda}{2}\int_{\Omega_+} |B(x,t)|^2\mm{d}y=\frac{\lambda}{2}\int_{\Omega_+}|\partial_{\bar{M}}\zeta(x,t)|^2\mm{d}y.$$

Denote by $\delta_{\bar{M}}(t)$ the variation of total magnetic energy from $0$ to $t$ moment. Then, we can use an integration by parts
and the boundary condition $\eta|_{\partial\Omega_+}=0$ to deduce
\begin{align}
\label{2017092117}
\delta_{\bar{M}}(t):=&\mathscr{E}_M^{\Omega_+}(t)-\mathscr{E}_M^{\Omega_+}(0) =\frac{\lambda }{2}\int_{\Omega_+}(|
\partial_{\bar{M}}  \zeta|^2 - |\bar{M}|^2)\mm{d}y \nonumber\\
=& {\lambda}\int_{\Omega_+}\partial_{\bar{M}}  \eta \cdot \bar{M} \mm{d}y+ \frac{\lambda }{2}\int_{\Omega_+}|\partial_{\bar{M}} \eta|^2\mm{d}y=\mathscr{V}_{\bar{M}}^{\Omega_+}(\eta(t))/2,
\end{align}
from which we immediately have the following physical conclusion.
\begin{concl}\label{physical:con31}
$\mathscr{V}_{\bar{M}}^{\Omega_+}(\eta(t))/2$ represents the variation of total magnetic energy of all the particles of the non-resistive MHD fluid
(defined on a periodic cell) deviating from their initial location at time $t$.
\end{concl}

From the relation \eqref{2017092117} and the estimate $\|\eta\|_{0,\Omega_+}^2\leqslant h^2\mathscr{V}_{\vec{n}}^{\Omega_+}(\eta)/\pi^2$
(see Remark \ref{201712161034} for a derivation), we find that the total magnetic energy increases, once the element lines are bent.
On the other hand, the bent element lines will be straightened by the magnetic tension from the figure on page \pageref{2017110011040}.
This dynamic behavior is very similar to that of the bent elastic string. Based on this dynamic behavior,
we can think that the non-resistive MHD fluid is made up of infinite elastic strings, and the magnetic energy is the elastic potential energy.

Next, we turn to the analysis of the physical meaning of $\mathscr{V}_{g\bar{\rho}' }^{\Omega_+} (\eta_3(t))$. During the development of RT instability,
the potential energy will be released by the interchange of heavier and lighter parts of the fluid. Hence, motivated by the physical meaning of $\mathscr{V}_{\bar{M}}^{\Omega_+}(\eta(t))$, we naturally guess that $\mathscr{V}_{g\bar{\rho}' }^{\Omega_+}(\eta_3(t))$ may be related to
the variation of gravitational potential energy from time $0$ to $t$.

 We denote by $\delta_{g\bar{\rho}'}(t)$ the variation of gravity potential energy. Recalling that initially only the initial velocity is perturbed,
  by the mass equation \eqref{0116}$_2$ we have
\begin{equation}
\label{20111041250}
\varrho=\varrho^0=\rho^0(\zeta^0)=\bar{\rho}(y_3),
\end{equation}
whence,
$$ \begin{aligned}
\delta_{g\bar{\rho}'}( t )=g\int_{\Omega_+}  (\rho-\bar{\rho})x_3\mm{d}x=g\int_{\Omega_+}  \bar{\rho} \zeta_3\mm{d}y-g\int_{\Omega_+}  \bar{\rho}x_3\mm{d}x=g\int_{\Omega_+}  \bar{\rho}\eta_3\mm{d}x, \end{aligned}  $$
where we have used the transform of Lagrange coordinates in the second equality.

To analyze the relation between $\mathscr{V}_{g\bar{\rho}' } (\eta_3(t)) $ and $\delta_{g\bar{\rho}}( t )$,  we next recall the mathematical
derivation of $\mathscr{V}_{g\bar{\rho}' }(\eta_3(t))$.

The relation \eqref{201706151213} in Lagrange coordinates reads as
\begin{equation}
\nabla_{\mathcal{A}} \bar{p}(\zeta_3)=-\bar{\rho}(\zeta_3)g  {e}_3,
\end{equation}
while using \eqref{201707161455} and \eqref{201706151104}, we calculate that
\begin{equation}
\label{201711122232}
  B\cdot \nabla_{\ml{A}} B=(\bar{M}\cdot\nabla )^2 {\eta}.
\end{equation}
Exploiting \eqref{20111041250}--\eqref{201711122232}, the momentum equation in Lagrange coordinates can be rewritten as
\begin{equation}
\label{201706151347}
\bar{\rho} u_t-\mu \Delta_{\ml{A}}u+\nabla_{\ml{A}}q=
\lambda  (\bar{M}\cdot\nabla )^2 {\eta}+G_{g\bar{\rho}} e_3,
\end{equation}
where $q:=q^*-\bar{p}(\zeta_3)$ and $G_{g\bar{\rho}}:=G_{g\bar{\rho}}(\eta_3,y_3):=g(\bar{\rho}(y_3+\eta_3(y,t))-\bar{\rho}(y_3))e_3$.
By the equation \eqref{201706151347}, we can regard $G_{g\bar{\rho}}$ as a force, which drives the growth of the RT instability in the fluid.

Let
$$\mathcal{N}_{g\bar{\rho}'} (\tau,y_3) :=g \int_{0}^{\tau} (\tau-z) \frac{\mm{d}^2}{\mm{d}z^2}\bar{\rho}(y_3 +z)\mm{d}z, $$
then
\begin{equation}  \label{201707172146}
G_{g\bar{\rho}}=g\bar{\rho}'\eta_3+\mathcal{N}_{g\bar{\rho}'}(\eta_3,y_3).
\end{equation}
Multiplying \eqref{201706151347} by $u$ in $L^2(\Omega_+)$, and making use integration by parts, the condition $\mm{div}_{\mathcal{A}}u=0$,
\eqref{0116}$_1$, the boundary condition \eqref{201712100806}, and the relation \eqref{201707172146}, we can deduce the following evolution
law for the variation of total energy of the non-resistive MHD fluid.
\begin{equation}
\label{201707182014}
\frac{\mm{d}\delta_{\mm{E}}^{\mm{N}}}{\mm{d}t} +\mu \|\nabla_{\mathcal{A}} u \|_{0,\Omega_+}^2=0,
\end{equation}
where
\begin{align}
&\delta_{\mm{E}}^{\mm{N}}(t):= \frac{1}{2}\int_{\Omega_+} \bar{\rho} |u(t)|^2\mm{d}y +\frac{1}{2}\mathscr{V}_{\bar{M}}^{\Omega_+}(\eta(t))-\mathscr{V}_{g \bar{\rho}'}^{* }(\eta_3(t)),\nonumber \\
& \mathscr{V}_{g\bar{\rho}'}^{* }(\eta_3(t))= \frac{1}{2}\mathscr{V}_{g\bar{\rho}'}^{\Omega_+}(\eta_3(t)) + \mathfrak{N}_{g\bar{\rho}}(\eta_3(t)),\quad \mathfrak{N}_{g\bar{\rho}} (\eta_3(t)):=
 \int_{\Omega_+}\int_0^{\eta_3(t)}\mathcal{N}_{g \bar{\rho}'} (\tau,y_3) \mm{d}\tau\mm{d}y.\nonumber
\end{align}

Noting that the first two integrals in $\delta_{\mm{E}}^{\mm{N}}$ represent the variations of kinetic and magnetic energies respectively,
one could guess that $-\mathscr{V}_{g\bar{\rho}}^*(\eta_3(t))$ may represent the variation of potential energy from time $0$ to $t$, i.e.,
\begin{equation}
\label{201707171716}
-\mathscr{V}_{g\bar{\rho}' }^{*}(\eta_3(t))=\delta_{g \bar{\rho}'}(t).
\end{equation}
In particular, if $\bar{\rho}'$ is a constant, \eqref{201707171716} reduces to
$$-\frac{1}{2}\mathscr{V}_{g\bar{\rho}' }^{\Omega_+}(\eta_3(t)) =\delta_{g\bar{\rho}' }(t).$$
Next we verify \eqref{201707171716}.

Let
\begin{equation*}
\mathscr{W}_{g\bar{\rho}'}(t):
=\int_{\Omega_+} \int^t_0 G_{g\bar{\rho}}(\eta_3(y,\tau),y_3  ) u_3(y,\tau)  \mm{d}\tau\mm{d}y .
\end{equation*}
The physical meaning of $\mathscr{W}_{g\bar{\rho}'}(t)$ will be discussed at the end of this section.
Recalling the relation \eqref{201707172146}, we have
$$\frac{\mm{d}}{\mm{d}t}\mathscr{W}_{g\bar{\rho}'}(t)=\int_{\Omega_+} G_{g\bar{\rho}}(\eta_3(y,t),y_3  ) u_3(y,t)  \mm{d}y
=\frac{\mm{d}}{\mm{d}t}\mathscr{V}_{g \bar{\rho}'}^{* }(t),$$
which, together with the fact that $\mathscr{W}_{g\bar{\rho}'}|_{t=0}=0$ and $\eta|_{t=0}=0$, implies
$$ \mathscr{W}_{g\bar{\rho}'}(t) = \mathscr{V}_{g \bar{\rho}'}^{* }(t).$$
Thus, to get \eqref{201707171716},  we only need to show
\begin{equation} \label{20170712113}
-\mathscr{W}_{g\bar{\rho}'}(t)=\delta_{g\bar{\rho}'}(t) . \end{equation}

Recalling the definition of $G_{g\bar{\rho}'}$ and the fact $\eta|_{t=0}=0$, we find that
\begin{align}
 \mathscr{W}_{g\bar{\rho}'}(t)= & g\int_{\Omega_+}\int_0^t (\bar{\rho}(y_3+\eta_3(y,\tau))-\bar{\rho}(y_3))u_3(y,\tau)\mm{d}\tau\mm{d}y\nonumber \\
= & g\int_{\Omega_+} \int_0^t \bar{\rho}(y_3+\eta_3(y,\tau))u_3(y,\tau)\mm{d}\tau\mm{d}y-g\int_{\Omega_+}\bar{\rho}
{\eta_3}\mm{d}y. \label{201708312033}
\end{align}
In addition, we have
$$
\begin{aligned}
&\frac{\mm{d}}{\mm{d}t}\int_{\Omega_+} \int_0^{t}  \bar{\rho}(y_3+\eta_3(y,\tau))u_3(y,\tau)\mm{d}\tau\mm{d}y=\int_{\Omega_+}\bar{\rho}(\zeta_3) u_3\mm{d}y
\\
&=\int_{\Omega_+}\bar{\rho}v_3\mm{d}x=\int_0^h\bar{\rho}\int_{ \Sigma_{x_3} } v_3\mm{d}x_{\mm{h}}\mm{d}x_3= \int_0^h\bar{\rho}\int_{\Omega_0^{x_3}}\mm{div}v \mm{d}x\mm{d}x_3=0,
\end{aligned}$$
where we have used the transform of Lagrangian coordinates and \eqref{zetanabla} in the second equality,
 integration by parts and the non-slip boundary condition for the velocity in the fourth equality,
 and the divergence-free condition in the last equality. Therefore, one concludes that
$$ \int_{\Omega_+} \int_0^t  \bar{\rho}(y_3+\eta_3(y,\tau))u_3(y,\tau)\mm{d}\tau\mm{d}y=0.$$
Substitution of the above identity into \eqref{201708312033} yields \eqref{20170712113}.
Consequently, one sees that $-\mathscr{V}_{g\bar{\rho}'}^{\Omega_+}(\eta(t))/2$ does not represent the variation of potential energy from time $0$ to $t$ except for $\bar{\rho}'$ being a constant.

However, $-\mathscr{V}_{g\bar{\rho}'}^{\Omega_+}(\eta(t))/2$ can approximately represent the variation of potential energy.
In fact, for any $w\in L^\infty(\Omega_+)$ satisfying
$$ w+y\in \Omega_+\mbox{ a.e. in }\Omega_+, $$
one can estimate that
\begin{align}
\label{201709062212}
\left|\mathfrak{N}_{g\bar{\rho}} (w)\right|\leqslant c \int_{\Omega_+}|w_3|^3 \mm{d}y\leqslant c\|w_3 \|_{L^\infty(\Omega_+)}\|w_3 \|_{0,\Omega_+}^2,
\end{align}
where the condition $\bar{\rho}\in C^2[0,h]$ has been used, and the constant $c$ only depends on $g$ and $\bar{\rho}'$.
In particular, if
 \begin{equation}  \label{201712131059}
\inf_{x_3\in (0,h)}\{\bar{\rho}'(x_3)\}>0,
\end{equation}
one further has
$$\left|\mathfrak{N}_{g\bar{\rho}} (w)\right| \leqslant c\|w_3 \|_{L^\infty(\Omega_+)}\mathscr{V}_{g\bar{\rho}'}^{\Omega_+}(w_3 ).$$
By virtue of the above estimate, for any given $t$, $\mathscr{V}_{g\bar{\rho}'}^{\Omega_+}(\eta_3(t))$ is an equivalent infinitesimal of
$\mathscr{V}_{g\bar{\rho}'}^*(\eta_3(t))$ as $\|\eta_3(t)\|_{L^\infty(\Omega_+)}\to 0$.

Summing up the above analysis, we arrive at the following physical conclusion:
\begin{concl}\label{physical:con32}
$-\mathscr{V}_{g\bar{\rho}'}^*(\eta(t))$ just represents the variation of potential energy of the MHD fluid (defined on a periodic cell)
from time $0$ to $t$.
Under the conditions $\bar{\rho}\in C^2[0,h]$ and \eqref{201712131059}, $\mathscr{V}_{g\bar{\rho}'}^{\Omega_+}(\eta_3(t))$ is approximately equal to $\mathscr{V}_{g\bar{\rho}'}^*(\eta_3(t))$ for any $t$ in the following sense:
$$\frac{2\mathscr{V}^*_{g\bar{\rho}'}(\eta_3(t))}{\mathscr{V}_{g\bar{\rho}'}^{\Omega_+}(\eta_3(t))}
=1+O(\|\eta_3(t)\|_{L^\infty(\Omega_+)}),$$
where $O(\|\eta_3(t)\|_{L^\infty(\Omega_+)})\to 0$ as $\|\eta_3(t)\|_{L^\infty(\Omega_+)}\to 0$.
In particular, if $\bar{\rho}'$ is a constant, then
 $-\mathscr{V}_{g\bar{\rho}'}(\eta_3(t))/2$ just represents the variation of potential energy from time $0$ to $t$.
\end{concl}

With Conclusions \ref{physical:con31}--\ref{physical:con32} in hand, we are in a position to analyze the physical meaning of the stability
and instability criteria for the {NMRT} problem.

It is well-known that the RT instability can be explained by the minimum potential energy principle, which is a fundamental concept used in physics,
chemistry, biology, and engineering, etc. It dictates that a structure or body shall deform or displace to a position that (locally) minimizes the total
 potential energy, with the lost potential energy being converted into kinetic energy (specifically heat).
 As aforementioned, the magnetic energy can be regarded  as the elastic potential energy, we thus call
$$\delta_{\mm{E}_{\mm{P}}}^{\mm{N}}(\eta(t)):=\frac{1}{2}\mathscr{V}_{\bar{M}}^{\Omega_+}(\eta)-\mathscr{V}_{g \bar{\rho}'}^{* }(\eta_3)$$
the variation of total potential energy, which depends on the displacement function of particles in the non-resistive MHD fluid.

Now we further denote the total potential energy in the rest state by $E_{\mm{P}}^{\mm{N},\mm{r}}$, then the total potential energy of the
non-resistive MHD fluid, denoted by $E_{\mm{P}}^{\mm{N}}(\eta(t))$ at time $t$, can be given by  $$E_{\mm{P}}^{\mm{N}}(\eta(t))=E_{\mm{P}}^{\mm{N},\mm{r}}+\delta_{E_{\mm{P}}}^{\mm{N}}(\eta(t)).$$
If one inserts the above relation into the evolution law of the total energy variation \eqref{201707182014}, one finds the following evolution law
for the total energy of the non-resistive MHD fluid:
\begin{equation}\label{201711031350}
\frac{\mm{d} }{\mm{d}t}\left( \frac{1}{2}\int_{\Omega_+} \bar{\rho} |u|^2\mm{d}y
+E_{\mm{P}}^{\mm{N}}(\eta) \right) +\mu \|\nabla_{\mathcal{A}} u \|_{0,\Omega_+}^2=0\footnote{
If $\bar{\rho}$ is a constant, $g=0$, $\mu=0$ and $\bar{M}:=(0,0,\bar{M}_3)$ in \eqref{201711031350}, then one has
$$ \int_{\mathbb{T}^2}\frac{\mm{d}}{\mm{d}t}E_{\mm{l}}(y_{\mm{h}},t)\mm{d}y_{\mm{h}} =0,$$
where $$E_{\mm{l}}(y_{\mm{h}},t):=\frac{1}{2}\left(\int_0^h |\partial_t \eta|^2\mm{d}y_3
+a^2\int_0^h\left| \partial_3 \eta\right|^2\mm{d}y_3\right)\;\mbox{ and }\; a=\sqrt{{\frac{\lambda }{\bar{\rho}}}}|\bar{M}_3|.$$
We easily see that $E_{\mm{l}}$ looks very similar to the energy of the one-dimensional (linear) wave equation
(or the chord vibration equation) with fixed boundary. By virtue of the figure on page \pageref{201711031343}, it is not surpring to see
 this similarity, since the bent element line will vibrate around its initial location. Hence $E_{\mm{l}}$ should repreasent the total energy
 of each bent element line, and the wave caused by this vibratation will propagate along the bent element line. This means that $a$ stands for
 the propagatation speed, and such a wave is called the Alfv\'en wave. The rigirous derivation of Alfv\'en wave based on the linearized MHD equation can be found in the classical textbook on MHD flows.}.
\end{equation}

In view of the above evolution law, we naturally believe that the minimum energy principle can be also used to explain the stability and instability
results in the {NMRT} problem. In other words, whether the potential energy in the rest state is minimal determines whether the {NMRT} problem is stable.
Hence, we guess that whether the potential energy in the rest state is minimal should be determined by the stability/instability conditions.
The following theorem indeed supports this conjecture.
\begin{thm}  \label{201707191444}
Let $\bar{\rho}\in C^2[0,h]$ satisfy the RT condition:
$$  \bar{\rho}'|_{x_3=x^0_{3}}>0\quad\mbox{ for some }\; x^0_{3}\in (0,h),$$
and $\bar{M}$ be a non-zero constant vector. Then, the following assertions hold.
\begin{enumerate}
\item[(1)] Under the instability condition
  \begin{equation}
\label{201709041500sdafsadfsa}
|\bar{M}_3|<{m}_{\mm{N}},
\end{equation}
  we have the non-minimal condition of total potential energy in the following sense:

For any given $k \geqslant 3$, there are a function $w\in C_\sigma^\infty(\Omega_+)$ and a constant $\varepsilon_0\in (0,1)$
(depending on $w$ and $k$), such that for any $\varepsilon\in (0,\varepsilon_0)$, there exist functions $\varpi_{\varepsilon,\mm{r}}$
(depending on $\varepsilon$) satisfying $\|\varpi_{\varepsilon,\mm{r}}\|_{k,\Omega_+}\leqslant c$, and
\begin{equation}
\label{201709060727sd}
\varpi:=\varepsilon w+ \varepsilon^2\varpi_{\varepsilon,\mm{r}} \in H^{k,1}_{0,*}(\Omega_+),\;\;\mbox{ and }\;\delta_{E_{\mm{P}}}^{\mm{N}}(\varpi)<0\mbox{ (i.e.  }E_{\mm{P}}^{\mm{N}}(\varpi)<E_{\mm{P}}^{\mm{N},\mm{r}}\mbox{)},
\end{equation}
where $c$ depends on $k$, $\|\varpi \|_{k,\Omega_+}$ and $\Omega_+$.
\item[(2)] Under the stability condition
 \begin{equation}
\label{201709041500sdaf}
|\bar{M}_3|>{m}_{\mm{N}},
\end{equation}
 we have the minimal condition of total potential energy, i.e., there is a constant $\varepsilon>0$ such that, for any non-zero $\varpi\in H_0^1(\Omega_+)$  satisfying $\|\varpi\|_{L^\infty(\Omega_+)}\leqslant \varepsilon$,
\begin{equation}
\label{201709060727}
\delta_{E_{\mm{P}}}^{\mm{N}}(\varpi)>0\mbox{ (i.e. }E_{\mm{P}}^{\mm{N}}(\varpi)>E_{\mm{P}}^{\mm{N},\mm{r}}).
\end{equation}
\end{enumerate}
\end{thm}
\begin{pf}
The proof  of Theorem \ref{201707191444} needs some new preliminary mathematical results, so we will provide the detailed proof in Sections \ref{201711241354} and \ref{201711241354f}.  \hfill $\Box$
\end{pf}

By Theorem \ref{201707191444}, we indeed have the following physical conclusion.
\begin{concl}\label{201711122037}
 Under the stability condition \eqref{201709041500sdaf} for $j=3$, the total potential energy in the rest state (defined on a periodic cell) is minimal
 in the sense of \eqref{201709060727}.
Under the instability condition \eqref{201709041500sdafsadfsa} for $j=3$,
the potential energy in the rest state is not minimal in the sense of \eqref{201709060727sd}.
\end{concl}

We have given the physical explanation of the stability/instability criteria for the {NMRT} problem. However, we can not judge whether
the {NMRT} problem is stable or not for the critical case $|\bar{M}_3|=m_{\mm{N}}$. In addition, under the instability
condition \eqref{201709041500sdafsadfsa}, we may expect that the perturbed MHD flow should tend to another rest (equilibrium) state,
the potential energy of which is minimal. Unfortunately, we can not show this mathematically.

Finally, we discuss the physical meaning of $\mathscr{W}_{g\bar{\rho}'}(t)$.
Since $\mathscr{W}_{g\bar{\rho}'}(t)$ is closely related to the force $G_{g\bar{\rho}}$, we first discuss the physical meaning of the force
$G_{g\bar{\rho}}$. To this end, we take a particle $Y$ labeled by $y$ in the MHD fluid to analyze the action of $G_{g\bar{\rho}}$
on the motion of the fluid without considering other forces. For the sake of simplicity, we assume that $\bar{\rho}'>0$. Then, from the relation $$G_{g\bar{\rho}}=g \int_{y_3}^{y_3+\eta_3(y,t)}\bar{\rho}'(s)\mm{d}s $$
we easily observe the following dynamic phenomena in the movement process of $Y$ from time $0$ to $t$:
\begin{itemize}
\item if $\eta_3(y,t)<0$ after perturbation, then the force $G_{g\bar{\rho}}$ drives the element point $Y$ sinking. This implies that $\eta_3(y,t)$
further decreases, and meanwhile $Y$ further goes down away from its initial location.
\item if $\eta_3(y,t)>0$ after perturbation, then the force $G_{g\bar{\rho}}$ drives the element point $Y$ up. This shows that $\eta_3(y,t)$
  further increases, and meanwhile $y$ further goes up away from its initial location.
\end{itemize}

The above analysis is consistent with the early growth stage of the RT instability under the force $G_{g\bar{\rho}}$.
Moreover, the above dynamic phenomena are caused by the pressure difference between the heavier and lighter fluids. Hence, we easily think
that $G_{g\bar{\rho}}$ represents the pressure difference. Moreover, $\mathscr{W}_{g\bar{\rho}'}(t)$ shall represent the total work done by the pressure difference $G_{g\bar{\rho}}$ to all particles of the fluid from time $0$ to $t$. We verify this fact below.

Consider a particle $Y$ labeled by $y$ deviating from its initial location to a new location, then the pressure difference
will do work to $Y$ in the motion process. To evaluate the work, we denote the location and the volume element of $Y$ at time $t$
by $\zeta(y,t)$ and $\mm{d}y$, respectively. Since the path $\Gamma$ of the particle $Y$ from its initial location to the new location $\zeta$
is continuously differentiable with respect to $(x,t)$, and the equation of $\Gamma$ can be given by
$$\Gamma: x=\zeta(y,s),\quad 0\leqslant s\leqslant t.$$
Thus, using the line integral of the second type and the relation $\mm{d}\zeta=u\mm{d}t$, we find that the work can be given by
\begin{equation*}
\int_{\Gamma}  \mm{d}y  G_{g\bar{\rho}}(x_3-y_3,y_3 ) e_3\cdot \mm{d}x=
\int^t_0 G_{g\bar{\rho}}(\eta_3(y,\tau),y_3 ) u_3(y,\tau) \mm{d}\tau\mm{d}y=:\mathscr{W}_{g\bar{\rho}'}^Y(y,t).
\end{equation*}
We integrate $\mathscr{W}_{g\bar{\rho}'}^Y(y,t)$ with respect to all particles of the fluid to see that $\mathscr{W}_{g\bar{\rho}'}(t)$ indeed
represents the total work done by the pressure difference $G_{g\bar{\rho}}$ to the all particles of the MHD fluid from time $0$ to $t$.
In addition, the relation \eqref{20170712113} tells us that the work done by the pressure difference comes from the release of potential energy.

\subsection{Case of the {SMRT} problem}

In this subsection we further extend the physical conclusions on the {NMRT} problem in the previous section to the {SMRT} problem.
We begin with introducing the mathematical model for the {SMRT} problem. Consider two distinct, immiscible, incompressible MHD fluids evolving
in a moving domain $\Omega(t)=\Omega_+(t)\cup\Omega_-(t)$ for time $t\geqslant 0$, where the upper fluid fills the upper domain
$$ \Omega_+(t):=\{x:=(x_{\mm{h}},x_3)\in \mathbb{R}^3~|~x_{\mm{h}}\in \mathbb{T}^2 ,\ d(x_{\mm{h}},t)< x_3<h\}, $$
and the lower fluid fills the lower domain
$$ \Omega_-(t):=\{x\in \mathbb{R}^3~|~x_{\mm{h}}\in \mathbb{T}^2 ,\ -l< x_3<d(x_{\mm{h}},t)\}. $$
We assume that $h$ and $l$ are given constants satisfying $h>- l$, but the internal surface function $d:=d(x_{\mm{h}}, t)$ is free and unknown.
The internal surface
$$\Sigma(t):=\{x\in \mathbb{R}^3~|~x_{\mm{h}}\in \mathbb{T}^2,\ x_3=d(x_{\mm{h}},t)\}$$
moves between the two MHD fluids, and $\{x_3 =-l \}$ and  $\{x_3 =h \}$ are the fixed lower and upper boundaries of $\Omega(t)$, respectively.

Now, we use the equations \eqref{comequations2151x}$_3$, \eqref{comequations2151x}$_4$ and \eqref{201706091542} with constant density
to describe the motion of stratified (uniform) incompressible MHD fluids (without resistivity), and add the subscript $+$, resp.
$_-$ to the notations of the known physical parameters, and other unknown functions in \eqref{comequations2151x}$_3$, \eqref{comequations2151x}$_4$ and
\eqref{201706091542} for the upper, resp. lower fluids. Thus, the motion equations of stratified incompressible MHD fluids driven by the uniform
gravitational field read  as follows.
\begin{equation}\label{0101f1}\left\{{\begin{array}{ll}
\rho_\pm \partial_t v_\pm+\rho_\pm v_\pm\cdot\nabla v_\pm+\mm{div}\mathcal{S}_\pm^g =0&\mbox{ in } \Omega_\pm(t),\\[1mm]
\partial_t M_\pm=M_\pm\cdot \nabla v_\pm-v_\pm\cdot\nabla M_\pm &\mbox{ in } \Omega_\pm(t),\\[1mm]
\mm{div} M_\pm=0,&\mbox{ in } \Omega_\pm(t),\end{array}}  \right.
\end{equation}
where $\rho_\pm$ are constants satisfying the RT jump  condition ${\rho}_+>\rho_-$, and
$\mathcal{S}_\pm^g : = \mathcal{S}^g(v_\pm,M_\pm,p^g_\pm):= p^g_\pm I-\mu_\pm \mathbb{D}(v_\pm)-\lambda M_\pm \otimes M_\pm$, $p^g_\pm:= p_\pm^*+\rho_\pm gx_3$
and
$\mathbb{D}(v_\pm)=\nabla v_\pm+\nabla v_\pm^{\mm{T}}$.

For two viscous MHD fluids meeting at a free boundary, the standard assumptions are that the velocity is continuous across the interface and that the jump in the normal stress is zero under ignoring the internal surface tension. This requires us to enforce the jump conditions
\begin{equation}\label{201612262131}
\llbracket v  \rrbracket=0 \quad \mbox{on }\Sigma(t),
\end{equation}
and
\begin{equation}\label{201711020923}
 \llbracket   \mathcal{S}^g \rrbracket\nu =G_{g\llbracket \rho \rrbracket }:=g \llbracket\rho\rrbracket d\nu\quad\mbox{on }\Sigma(t),
\end{equation}
where $\nu$ represents the unit out normal vector of $\Omega_-(t)$, $\llbracket f \rrbracket:=f_+|_{\Sigma(t)} -f_-|_{\Sigma(t)}$
denotes the interfacial jump, and $f_\pm|_{\Sigma(t)}$ are the traces of the functions $f_\pm$ on $\Sigma(t)$.
We will also enforce the condition that the fluid velocity vanishes at the upper and lower fixed boundaries, i.e.,
\begin{equation}\label{201612262117}
v=0 \quad \mbox{ on }\partial\Omega_{-l}^h.
\end{equation}
We also call $G_{g\llbracket \rho \rrbracket }$ the pressure difference caused by gravity, since, similarly to $G_{g\bar{\rho}'}$, it drives
the growth of the RT instability by acting on the particles on the interface. At the end of this section, we will further discuss the behavior of
$G_{g\llbracket \rho \rrbracket }$.

To simplify the formulation of \eqref{0101f1},   we introduce the indicator function $\chi$ and denote
 \begin{align}
& \mu:=\mu_+\chi_{\Omega_+(t)} +\mu_-\chi_{\Omega_-(t)},\;\; \rho:=\rho_+\chi_{\Omega_+(t)} +\rho_-\chi_{\Omega_-(t)},\;\; v:=v_+\chi_{\Omega_+(t)} +v_-\chi_{\Omega_-(t)},\nonumber \\
& M:=M_+\chi_{\Omega_+(t)} +M_-\chi_{\Omega_-(t)},\;\; p^*:=p_+^*\chi_{\Omega_+(t)} +p_-^*\chi_{\Omega_-(t)},\;\;
\mathcal{S}^g:= \mathcal{S}_+^g\chi_{\Omega_+(t)} + \mathcal{S}_-^g\chi_{\Omega_-(t)}  \nonumber
\end{align}
to arrive at
\begin{equation}\label{201611091004}\left\{{\begin{array}{ll}
\rho v_t+\rho v\cdot\nabla v+\mm{div}\mathcal{S}^g=0&\mbox{ in } \Omega(t),\\[1mm]
 M_t=M\cdot \nabla v-v\cdot\nabla M &  \mbox{ in } \Omega (t),\\
\mm{div }M=0&  \mbox{ in } \Omega (t). \end{array}}  \right.
\end{equation}
Moreover, by virtue of the boundary condition \eqref{201612262131}, the internal surface function is defined by $v$, i.e.,
\begin{equation}\label{201612262217}
\partial_t d +v_1 \partial_1d +v_2 \partial_2d =v_3 \mbox{ on }\Sigma(t).
\end{equation}

Finally, we impose the initial condition for $(v, M,d)$:
\begin{equation}
\label{201612262216n}
(v, M )|_{t=0}:=( v^0,M^0 )\;\;\mbox{ in }\; \Omega_{-l}^h\setminus \Sigma(0)\mbox{ and }
 d |_{t=0}= d^0 \mbox{ on } \mathbb{T}^2,
\end{equation}
where $\Sigma(0)=\{x\in \mathbb{R}^3~|~x_{\mm{h}}\in \mathbb{T}^2,\ x_3=d^0\}$.

Then \eqref{201612262131}--\eqref{201612262216n} constitute an initial boundary value problem of incompressible stratified
MHD fluids with a free interface, which we call the {SMF} model for simplicity.

Now, let us construct a rest state of the {SMF} model to be studied. Without loss of generality, assume the interface in the rest state is $\{x_3=0\}$.
 Let $\bar{M}$ be a uniform magnetic field in $\Omega $, and $\bar{p}^*_\pm$  satisfy
 \begin{equation*}
\left\{\begin{array}{ll}
\nabla \bar{p}_\pm^*  =0 &\mbox{ in } \Omega_\pm,\\[1mm]
     \llbracket   \bar{p}^*   \rrbracket  =0&\mbox{ on }\Sigma.
  \end{array}\right.
\end{equation*}
Then ${r_{\mm{S}}}:=(u=0,\bar{M},d=0,\bar{p}^*)$ is called the rest state of the {SMF} model.
The problem whether the rest state ${r_{\mm{S}}}$  is stable or unstable is called the {SMRT} problem.

In \cite{JFJSWWWOA}, Jiang, et.al. showed that the value $\sqrt{2 g\pi \llbracket \rho \rrbracket /\sigma}$ is a strength-threshold
of the impressed field $\bar{M}=(0,0,\bar{M}_3)^{\mm{T}}$ for stability/instability of the linearized {SMRT} problem with $h=1$ and $l=1$.
Then, Wang \cite{WYTIVNMI} further established the existence of an asymptotically stable solution for the (nonlinear) {SMRT} problem defined on $\Omega':=\mathbb{R}^2\times (-l,m)$
under the (asymptotical) stability condition
\begin{equation*}
|\bar{M}_3|>m_{\mm{S}}'
\end{equation*}
with
\begin{equation*}
m_{\mm{S}}':=\sqrt{\sup_{ w\in H_{\sigma}^1(\Omega')} \frac{\mathscr{V}_{g\llbracket \rho \rrbracket}' (w_3)}{\mathscr{V}_{e_3}^{\Omega'}(w)}},
\qquad \mathscr{V}_{g\llbracket \rho \rrbracket}'(w_3):= g\llbracket \rho \rrbracket \|w_3^2(\cdot,0)\|_{L^2(\mathbb{R}^2)}^2.
\end{equation*}
Moreover, Wang further gave that
\begin{equation*}
m_{\mm{S}}'=2\sqrt{\frac{ g\pi \llbracket \rho \rrbracket}{\sigma(h^{-1}+l^{-1})}}.
\end{equation*}
Of course, Wang's stability result also holds for the domain $\Omega_{-l}^h$.

The above stability result tells us that under the stability condition, the RT instability can be inhibited by a magnetic field. Of course, the magnetic inhibition mechanism in the {NMRT} problem can be also used to explain the above stability result.
Moreover, we believe that the stability condition has physical meaning as in the case of the NMRT problem.

In the following, we give the physical meaning of the stability condition
\begin{equation}\label{201707271549ads}
|\bar{M}_3|>{m_{\mm{S}}},
\end{equation}
  and the instability condition
\begin{equation} \label{201707271549}
|\bar{M}_3|<{m_{\mm{S}}}
\end{equation}
in Lagrangian coordinates with proper regularity assumption on the solutions,
where we have defined that
\begin{equation*}
m_{\mm{S}}:=\sqrt{\sup_{ w\in H_{\sigma}^1(\Omega_{-l}^h)} \frac{\mathscr{V}_{g\llbracket \rho \rrbracket} (w_3)}{\mathscr{V}_{e_3}^{\Omega }(w)}},
\qquad \mathscr{V}_{g\llbracket \rho \rrbracket}(w_3):= g\llbracket \rho \rrbracket |w_3^2(\cdot,0)|_{0}^2.
\end{equation*}
 In particular, we pay attention to the detailed derivation of
the physical meaning of $-\mathscr{V}_{g\llbracket \rho \rrbracket} (\eta_3(t))$, where some mathematical techniques are required. We mention that our analysis results also hold for the domain $\Omega'$.

Now, let us disturb the rest state ${r_{\mm{S}}}$ in the velocity by $v^0$, and assume that the SMHD model describing the motion of the MHD fluid
after perturbation defines a classical solution $(v,M,d,p^*)$ defined on $\overline{Q^{t,T}_+}\cup\overline{Q^{t,T}_-}$,
where $Q^{t,T}_\pm:=\{(x,t)~|~t\in  [0,T],\ x\in \Omega_\pm (t)\}$ for some $T>0$.
 To make some integrals involving $v$ and $p^*$ sense, we assume that
$$v_\pm\in C^1(\overline{Q^{t,T}_\pm })\mbox{ and }p^*_\pm\in C^0(\overline{ Q^{t,T}_\pm }).$$
We further assume that $\zeta_\pm$ is the solution of
\begin{equation*}
\left\{
\begin{array}{ll}
\partial_t \zeta_\pm (y,t)=v_\pm(\zeta_\pm(y,t),t), & y\in\overline{\Omega_\pm}, \\
\zeta_\pm(y,0)=y, & y\in \overline{\Omega_\pm}
\end{array}    \right.
\end{equation*}
with the regularity $\zeta_\pm(\cdot,t)\in C^2(Q_\pm^T)$ and $\zeta_\pm \in C^1(\overline{Q^T_\pm})$,
where $Q_\pm^T:=\Omega_\pm\times [0,T]$.
We further assume that for each $t\in [0, T]$,
\begin{equation}
\label{201712101359}
\zeta_\pm(t):\overline{\Omega_\pm}\to \overline{\Omega_\pm(t)}\;\;\mbox{  are reversible},
 \end{equation}
and $\Sigma(t)=\zeta_\pm(\Sigma,t)$. Moreover, by virtue of the non-slip boundary condition \eqref{201612262117} and the continuity of
the velocity across $\Sigma$, $\zeta$ satisfies
\begin{equation*}
y=\zeta_\pm(y,t)\;\mbox{ on }\;\Sigma_\pm \;\; \mbox{ and } \;
\llbracket   \zeta\rrbracket=0\mbox{ on } \Sigma.
\end{equation*}

From now on, we denote
\begin{equation}
\label{201711141855}
\zeta=\zeta_+\chi_{\Omega_+}+\zeta_-\chi_{\Omega_-},\;\;\zeta_0=\zeta_+^0\chi_{\Omega_+}+\zeta_-^0\chi_{\Omega_-}\;\mbox{ and }\; \eta=\zeta-y.
\end{equation}
Obviously, $\zeta$ still enjoys the properties \eqref{201707161455}--\eqref{diverelation}.
Denoting
$$ ( u,B,q)(y,t)=(v,M,p^*)(\zeta(y,t),t),$$
  we obtain
$$B=\partial_{\bar{M}} \zeta\quad\mbox{and}\quad
 \mm{div}_{\mathcal{A}}( B \otimes B)= \partial_{\bar{M}}^2\eta,$$
where $\mathcal{A}$ is defined by \eqref{201712091617} with $\zeta$ given by \eqref{201711141855}.
Moreover,
$$B\otimes B \mathcal{A}e_3=\partial_{\bar{M}} \zeta \cdot ( \mathcal{A}e_3) \partial_{\bar{M}} \zeta=\bar{M}_3\partial_{\bar{M}} \zeta  .$$
Consequently, we derive from the SMHD model that $(\eta, u,B,q)$ satisfies the following initial-boundary value problem with an internal interface:
 \begin{equation}
\label{201611091004nn}
\left\{\begin{array}{ll}
\zeta_t=u&\mbox{ in } \Omega ,\\[1mm]
\rho u_t+\mm{div}_{\mathcal{A}}(qI-\mu \mathbb{D}_{\mathcal{A}}(u))=\lambda \partial_{\bar{M}}^2\eta&\mbox{ in } \Omega ,\\[1mm]
 \llbracket qI -\mu \mathbb{D}_{\mathcal{A}}(u)\rrbracket  {\mathcal{A}e_3} -
 \lambda \bar{M}_3 \llbracket\partial_{\bar{M}} \eta\rrbracket \\
 \quad =g \llbracket \rho \rrbracket \eta_3 {\mathcal{A}e_3}  &\mbox{ on }\Sigma,\\[2mm]
\llbracket  \eta\rrbracket=0& \mbox{ on } \Sigma,\\
(u,\eta)=0 & \mbox{ on }\Sigma_{-l}\cup \Sigma_h,\\
 (u,\eta)=(u^0,0) & \mbox{ at }t=0,
\end{array}\right.
\end{equation}
where $\mathbb{D}_{\mathcal{A}}(u):=\nabla_{\mathcal{A}} u+\nabla_{\mathcal{A}} u^{\mm{T}}$.

Since $\zeta(\cdot,t)$ is continuous across $\Sigma$, one has
\begin{equation}    \label{201712101415}
\llbracket  \mathcal{A}e_3\rrbracket=0\mbox{ on } \Sigma.
\end{equation}

Now, multiplying \eqref{201611091004nn}$_2$ by $u$ in $L^2(\Omega)$, and employing integration by parts, and the boundary conditions \eqref{201611091004nn}$_3$--\eqref{201611091004nn}$_5$ and \eqref{201712101415}, we infer
 \begin{align}\label{2017111603}
\frac{\mm{d}\delta_{E}^{\mm{S}}(t)}{\mm{d}t} +
\frac{\mu}{2}\|\mathbb{D}_{\mathcal{A}}(u(t))\|_{0,\Omega}^2=0,
\end{align}
where
\begin{align}
&\delta_{E}^{\mm{S}}(t):=\frac{1}{2}\|\sqrt{\rho} u(t)\|^2_{0,\Omega}+\delta_{E_\mm{P}}^{\mm{S}}(t),\quad \delta_{E_\mm{P}}^\mm{S}(t):=\frac{1}{2}\mathscr{V}_{\bar{M}}^{\Omega}(\eta(t))
 -\mathscr{V}_{g\llbracket \rho \rrbracket}^*(t),\nonumber \\
&\mathscr{V}_{g\llbracket \rho \rrbracket}^{*}(t):=
\frac{1}{2}\mathscr{V}_{g\llbracket \rho \rrbracket}(\eta_3(t))+ \mathfrak{N}_{g\llbracket \rho \rrbracket}(t)\nonumber ,\\
&\mathfrak{N}_{g\llbracket \rho \rrbracket}(t)=g{ \llbracket \rho \rrbracket }\int_{\Sigma}\int_0^t\eta_3(y,\tau)\tilde{\mathcal{A}}(y,\tau)e_3 \cdot u(y,\tau)\mm{d}\tau\mm{d}y_{\mm{h}}.\nonumber
\end{align}
Then, inspired by Conclusions \ref{physical:con31} and \ref{physical:con32}, we have the following physical conclusions.
\begin{concl}    \label{201707201630n}
\begin{enumerate}
\item[\quad (1)] $\mathscr{V}_{\bar{M}}^{\Omega}(\eta(t))/2$ represents the variation of magnetic energy of the stratified MHD fluid
without resistivity (defined on a periodic cell) deviating from its rest state ${r_{\mm{S}}}$ at time $t$;
\item[\quad (2)] $-\mathscr{V}^*_{g\llbracket \rho \rrbracket}(t)$ represents the variation of gravity potential energy  of the stratified MHD fluid
 deviating from its rest state ${r_{\mm{S}}}$ at time $t$.
\item[\quad (3)] For any given $t$, $\mathscr{V}_{g\llbracket\rho\rrbracket}(\eta_3(t))/2$ is approximately equal to
  $\mathscr{V}^*_{g\llbracket\rho\rrbracket}(t)$ in the following sense:
$$\frac{2\mathscr{V}^*_{g\llbracket\rho\rrbracket}(t)}{\mathscr{V}_{g\llbracket\rho\rrbracket}(\eta_3(t))}=1+O(f(t)),$$
where $f(t):=\|\nabla_{\mm{h}}\eta_{\mm{h}}(t)\|_{L^\infty(\Sigma)}(1+\| \nabla_{\mm{h}}\eta_{\mm{h}}(t) \|_{L^\infty(\Sigma)})$,
and $O(f(t))\to 0$ as $f(t)\to 0$.
\end{enumerate}
\end{concl}
\par \qquad
\\[-1.2em]D{\small{ERIVATION}}.
(1) The first conclusion obviously holds by following the arguments in the derivation of \eqref{physical:con31}.

(2) The derivation of the second conclusion requires more mathematical techniques. Let $\delta_{g\llbracket\rho\rrbracket}(t)$ denote
the variation of gravity potential energy from time $0$ to $t$. Then, we evaluate that
\begin{align}
\delta_{g\llbracket \rho \rrbracket}(t)=
 &g\left(\int_{\Omega(t)} \rho x_3 \mm{d}x-\int_{\Omega } \rho x_3 \mm{d}x\right)= g\left(\int_{\Omega }
 \rho \zeta_3 \mm{d}y-\int_{\Omega } \rho x_3 \mm{d}x\right)= g\int_{\Omega } \rho \eta_3 \mm{d}y\nonumber\\
=& g\int_{-l}^h \rho\int_{\Omega_{-l}^\tau} \mm{div}\eta \mm{d}y\mm{d}\tau=:\tilde{\delta}_{g\llbracket \rho \rrbracket}(t).\nonumber
 \end{align}
Thus, to establish the second conclusion, it suffices to show
\begin{equation}
\label{201709091358}
\tilde{\delta}_{g\llbracket \rho \rrbracket}(t)=-\mathscr{V}_{g\llbracket \rho \rrbracket}^{*}(t).
\end{equation}

Recalling $J=1$, we have that
\begin{align}
\mm{div}\eta=&\partial_1\eta_2\partial_2 \eta_1+ \partial_2 \eta_3\partial_3\eta_2+\partial_ 3\eta_1\partial_1\eta_3
-\partial_1\eta_1\partial_2 \eta_2-
\partial_1 \eta_1\partial_3\eta_3
\nonumber \\
&-\partial_2\eta_2\partial_3\eta_3+ \partial_1\eta_1(\partial_2\eta_3\partial_3\eta_2 -\partial_2\eta_2\partial_3\eta_3) \nonumber  \\
& + \partial_2\eta_1(\partial_1\eta_2\partial_3\eta_3 -\partial_1\eta_3\partial_3\eta_2)
+\partial_3\eta_1(\partial_1\eta_3\partial_2\eta_2 -\partial_1\eta_2\partial_2\eta_3).
\label{201707222033}
\end{align}
Integrating by parts, we further get
$$\int_{\Omega_{-l}^\tau} \mm{div}\eta \mm{d}y  = g\int_{\Sigma_\tau}( \eta_1(\partial_1\eta_3\partial_2\eta_2
-\partial_1\eta_2\partial_2\eta_3)-\eta_3\partial_1 \eta_1-\eta_3\partial_2\eta_2 )\mm{d}y_{\mm{h}}.$$
So, we obtain
\begin{align}
\label{201707231440}
\tilde{\delta}_{g\llbracket \rho \rrbracket}(t)=& g\int_{\Omega } \rho(
\eta_1(\partial_1\eta_3 \partial_2\eta_2
-\partial_1\eta_2\partial_2\eta_3)-\eta_3\partial_1\eta_1-\eta_3\partial_2\eta_2)\mm{d}y \nonumber \\
=& g\int_{\Omega } \rho \partial_3\eta_3\eta_3 \mm{d}y-I_{1}(t) =-\frac{1}{2}\mathscr{V}_{g\llbracket \rho \rrbracket}(\eta)-I_{1}(t),
\end{align}
where
$$I_{1}(t):=g\int_{\Omega } \rho(\mm{div}\eta\eta_3-\eta_1(\partial_1\eta_3 \partial_2\eta_2 -\partial_1\eta_2\partial_2\eta_3))\mm{d}y.$$

Obviously, to show \eqref{201709091358}, we have to establish
\begin{equation*}
I_1(t)=\mathfrak{N}_{g\llbracket \rho \rrbracket}(t).
\end{equation*}
Since $\eta|_{t=0}=0$ and $\mathfrak{N}_{g\llbracket\rho\rrbracket}(0)=0$, to get the above relation, it suffices to establish that
\begin{equation}\label{201707231437}
\frac{\mm{d}I_1(t)}{\mm{d}t}=\frac{\mm{d} \mathfrak{N}_{g\llbracket \rho \rrbracket}(t)}{\mm{d}t}.
\end{equation}

Employing the second identity in \eqref{diverelation} and partial integrations, we arrive at
\begin{align}
\frac{\mm{d}I_1(t)}{\mm{d}t}= & g\int_{\Omega } \rho(\mm{div}\eta u_3-\partial_t(\eta_1(\partial_1\eta_3 \partial_2\eta_2
-\partial_1\eta_2\partial_2\eta_3)))\mm{d}y+ g\int_{\Omega } \rho \mm{div}u \eta_3\mm{d}y \nonumber \\
= & g\int_{\Omega } \rho(\mm{div}\eta u_3-\partial_t(\eta_1(\partial_1\eta_3 \partial_2\eta_2
-\partial_1\eta_2\partial_2\eta_3)))\mm{d}y \nonumber  \\
& - g\int_{\Omega } \rho\mm{div}(u^{\mm{T}}\tilde{\mathcal{A}}) \eta_3\mm{d}y=I_2(t)
+\frac{\mm{d} \mathfrak{N}_{g\llbracket \rho \rrbracket}(t)}{\mm{d}t},\nonumber
\end{align}
where
$$I_2(t):=g\int_{\Omega } \rho\big( \mm{div}\eta u_3+(\tilde{\mathcal{A}}\nabla \eta_3)\cdot u-\partial_t(\eta_1(\partial_1\eta_3 \partial_2\eta_2-\partial_1\eta_2\partial_2\eta_3)) \big) \mm{d}y. $$
Next, we shall verify the following identity in order to get \eqref{201707231437}:
\begin{equation}   \label{201707231435}  I_2(t)=0.   \end{equation}

By a straightforward computation, we find that
$$  \begin{aligned}
&\int_{\Omega } \rho \tilde{\mathcal{A}}\nabla \eta_3\cdot u \mm{d}y=
\int_{\Omega } \rho( (A_{ij}^*)_{3\times 3 }-I)\nabla \eta_3\cdot u \mm{d}y\\
&=\int_\Omega \rho (u_1(\partial_2\eta_2\partial_1\eta_3-
\partial_1\eta_2\partial_2\eta_3)+u_2(\partial_1\eta_1\partial_2\eta_3- \partial_2\eta_1\partial_1\eta_3) \\
&\quad +u_3((\partial_2\eta_1\partial_3\eta_2-\partial_2\eta_2\partial_3\eta_1-\partial_3\eta_1)
\partial_1\eta_3+(\partial_1\eta_2\partial_3\eta_1-\partial_1\eta_1\partial_3\eta_2 -\partial_3\eta_2) \partial_2\eta_3 \\
&\quad +(\partial_1\eta_1\partial_2\eta_2-\partial_1\eta_2\partial_2\eta_1
+\partial_1\eta_1+\partial_2\eta_2) \partial_3\eta_3))\mm{d}y,
\end{aligned}$$
where $A^{*}_{ij}$ is the algebraic complement minor of the $(i,j)$-th entry $\partial_j \zeta_i$.
Thus, we employ \eqref{201707222033} and integration by parts to deduce
$$ \begin{aligned}
&\int_{\Omega } \rho( \mm{div}\eta u_3+(\tilde{\mathcal{A}}\nabla \eta_3)\cdot u )\mm{d}y\\
= &\int_{\Omega } \rho(u_1(\partial_2\eta_2\partial_1\eta_3 -
\partial_1\eta_2\partial_2\eta_3 )   +u_2(\partial_1\eta_1\partial_2\eta_3-  \partial_2\eta_1\partial_1\eta_3)
 +u_3(\partial_1\eta_2\partial_2 \eta_1-\partial_1\eta_1\partial_2 \eta_2)) \mm{d}y\\
=& \int_{\Omega } \rho \partial_t(\eta_1(\partial_1\eta_3 \partial_2\eta_2 -\partial_1\eta_2\partial_2\eta_3)) \mm{d}y .
\end{aligned} $$
Hence, \eqref{201707231435} holds, and one gets the desired conclusion:
\begin{equation}   \label{201712102027}
 -\mathscr{V}_{g\llbracket \rho \rrbracket}^{*}(t)=\delta_{g\llbracket \rho \rrbracket}(t).
\end{equation}

(3) Let
\begin{equation*}
I_3=\frac{  g \llbracket \rho \rrbracket  } {2} \int_{\Sigma }\eta^2_3(\partial_1 \eta_1\partial_2 \eta_2
+\partial_1 \eta_1+\partial_2 \eta_2 -\partial_2 \eta_1\partial_1 \eta_2 )\mm{d}y_{\mm{h}},
\end{equation*}
then one can deduce that
 $$   \begin{aligned} \frac{\mm{d} \mathfrak{N}_{g\llbracket \rho \rrbracket}(t)}{\mm{d}t}
=&  g \llbracket \rho \rrbracket \int_{{\Sigma}}\eta_3(( \partial_1\eta_2\partial_2\eta_3 -
\partial_1\eta_3\partial_2\zeta_2)u_1+( \partial_1\eta_3\partial_2\eta_1 - \partial_1\zeta_1\partial_2\eta_3)u_2\\
&\qquad \qquad +( \partial_1\zeta_1\partial_2\zeta_2 - \partial_1\eta_2\partial_2\eta_1-1) u_3 )\mm{d}y_{\mm{h}}  \\
=  &g \llbracket \rho \rrbracket  \int_{\Sigma } \bigg(\eta_3(\partial_1 \eta_1\partial_2 \eta_2
+\partial_1 \eta_1+\partial_2 \eta_2  -\partial_2 \eta_1\partial_1 \eta_2 )u_3\\
&\qquad\quad +\frac{1}{2} \eta_3^2(\partial_1 u_1\partial_2 \eta_2 +\partial_1 \eta_1\partial_2 u_2+\partial_1 u_1
+ \partial_2 u_2 -\partial_2 u_1\partial_1 \eta_2-\partial_2 \eta_1\partial_1 u_2 )\bigg)\mm{d}y_{\mm{h}} \\
= &\frac{\mm{d}}{\mm{d}t}I_3(t),\end{aligned} $$
which gives
\begin{equation}   \label{201711142213}
I_3(t)= \mathfrak{N}_{g\llbracket \rho \rrbracket}(t).
\end{equation}
Recalling the definition of $I_3(t)$, we infer that
\begin{equation}
\label{201711262224567894}
\left|\mathfrak{N}_{g\llbracket\rho\rrbracket}(t)\right|\leqslant\|\nabla_{\mm{h}}\eta_{\mm{h}}(t)\|_{L^\infty(\Sigma)}(1+\|\nabla_{\mm{h}} \eta_{\mm{h}}(t)\|_{L^\infty(\Sigma)})\mathscr{V}_{g\llbracket\rho\rrbracket}(\eta(t)).
\end{equation}
Thus, the desired conclusion follows immediately. \hfill$\Box$

In view of the relation \eqref{201711142213}, from now on, we renew to define $\mathfrak{N}_{g\llbracket \rho \rrbracket}(t)$ as follows.
\begin{align}
\mathfrak{N}_{g\llbracket \rho \rrbracket}(t)&:=\mathfrak{N}_{g\llbracket \rho \rrbracket}(\eta(t))\nonumber \\
&:=\frac{   \llbracket \rho \rrbracket  } {2} \int_{\Sigma}\eta^2_3(t)(\partial_1 \eta_1(t)\partial_2 \eta_2(t)
+\partial_1 \eta_1(t)+\partial_2 \eta_2(t)
-\partial_2 \eta_1(t)\partial_1 \eta_2(t) )\mm{d}y_{\mm{h}}.
\label{201711262224}
\end{align}
Similarly to Theorem \ref{201707191444}, we have
\begin{thm}   \label{201707191444n}
Let $\mathfrak{N}_{g\llbracket\rho\rrbracket}(t)\in\delta_{E_{\mm{P}}}^{\mm{S}}(\varpi)$ be defined by $\eqref{201711262224}$,
then the following assertions hold.
\begin{enumerate}
\item[(1)] Under the instability condition \eqref{201707271549}, we have the non-minimal condition of total potential energy in the following sense:

For any $k\geqslant 3$, there are a function $w\in C_\sigma^\infty(\Omega)$ and a constant $\varepsilon_0\in (0,1)$ (depending on $w$ and $k$), such that
for any $\varepsilon\in (0, \varepsilon_0)$, there exist functions $\varpi_{\varepsilon,\mm{r}}$ (depending on $\varepsilon$) satisfying
\begin{equation}    \label{201709060727sdsfads}
\varpi:=\varepsilon w+ \varepsilon^2\varpi_{\varepsilon,\mm{r}}  \in H^{k,1}_{0,*}(\Omega),\quad \|\varpi_{\varepsilon,\mm{r}}\|_{k,\Omega}
\leqslant c \mbox{ and }\delta_{E_{\mm{P}}}^{\mm{S}}(\varpi)<0,
\end{equation}
where $c$ depends on $k$, $\|w\|_{k,\Omega}$ and $\Omega$.

\item[(2)] Under the stability condition \eqref{201707271549ads}, we have the minimal condition of total potential energy, i.e., there is a constant $\varepsilon>0$, such that for any
  non-zero  $\varpi\in H_{0}^1(\Omega)$ satisfying $\|\nabla_{\mm{h}}\varpi_{\mm{h}}\|_{C^0(\mathbb{T}^2)}\leqslant \varepsilon$,
\begin{equation}   \label{201709060727sdfsaf}
\delta_{E_{\mm{P}}}^{\mm{S}}(\varpi)>0.
\end{equation}
\end{enumerate}
\end{thm}
\begin{pf}
We shall mention how to prove Theorem \ref{201707191444n} in Section \ref{201711241354f1907}.
\hfill $\Box$
\end{pf}

By Theorem \ref{201707191444n}, we immediately have the following physical explanation:
\begin{concl} Under the stability condition \eqref{201707271549ads}, the total potential energy in the rest state ${r_{\mm{S}}}$
is minimal in the sense of \eqref{201709060727sdfsaf}. Under the instability condition \eqref{201707271549}, the total potential energy
in the rest state ${r_{\mm{S}}}$ is not minimal in the sense of \eqref{201709060727sdsfads}.
\end{concl}

Next, we explain how to find the relation \eqref{201711142213} from the physical point of view.
First, $\delta_{g\llbracket \rho \rrbracket}(t)$ can be expressed by $d$, i.e.,
 $$ \begin{aligned}
\delta_{g\llbracket \rho \rrbracket}(t)= & g\left( \int_{\Omega(t)} \rho x_3 \mm{d}x-\int_{\Omega} \rho x_3 \mm{d}x\right)\\
= &g\left(\rho_+\int_{\mathbb{T}^2}\int_d^h x_3 \mm{d}x_3  \mm{d}x_{\mm{h}} +\rho_-\int_{\mathbb{T}^2}\int_{-l}^d x_3 \mm{d}x_{\mm{h}} \right)\\
& -2\pi^2 L_1L_2\left(\rho_+h^2-\rho_-l^2 \right)= -\frac{ g\llbracket \rho \rrbracket }{2}\int_{\mathbb{T}^2}d^2\mm{d}x_{\mm{h}}.
 \end{aligned}$$

By \eqref{201712101359}, the function $\zeta(\cdot,0,t):\mathbb{T}^2\mapsto \Sigma(t)$ is reversible for any $t\in [0,T]$.
Then, for any given $(x_1,x_2,d(x_{\mm{h}},t))$ on $\Sigma(t)$, there is a point $(y_{\mm{h}},0)$, such that
$$x_1=\zeta_1(y_{\mm{h}},0,t),\quad x_2=\zeta_2(y_{\mm{h}},0,t)\mbox{ and }d(x_{\mm{h}},t)=\zeta_3(y_{\mm{h}},0,t),$$
which yields
$$d(\zeta_1(y_{\mm{h}},0,t), \zeta_2(y_{\mm{h}},0,t),t)=\zeta_3(y_{\mm{h}},0,t).$$
Thus, if $\eta$ is suitably small, then we can use the above change of variables to formally have
$$ \begin{aligned}
\delta_{g\llbracket \rho \rrbracket}(t):= &-\frac{  \llbracket \rho \rrbracket } {2} \int_{\Sigma}\zeta^2_3(\partial_1 \zeta_1\partial_2 \zeta_2
-\partial_2 \zeta_1\partial_1 \zeta_2 )\mm{d}y_{\mm{h}}= - \mathscr{V}_{g\llbracket \rho \rrbracket}(\eta_3(t))/2- I_3(t).
 \end{aligned}$$
Recalling the definition of $\mathscr{V}_{g\llbracket \rho \rrbracket}^*(t)$ and the relation \eqref{201712102027}, we easily obtain \eqref{201711142213}.

Finally we derive some additional physical results involving the pressure difference $G_{g\llbracket\rho\rrbracket }$.
We consider some part of the upper heavier fluid at time $t$, which sinks below the interface ${\Sigma}(t)$. Assume that the
domain $\mathfrak{D}_{\mm{s}}$ occupied by the sinking part is just a bounded connected domain that is bounded by the plane $\{x_3=0\}$
and the surface $\Sigma(t)$. Then, the total pressure difference $G_{\mathfrak{D}_{\mm{s}}}$ acting on the lower surface $\mathfrak{D}_{\mm{s}}$
can be given by the following integral formula:
$$G_{\mathfrak{D}_{\mm{s}}}= \int_{\overline{\mathfrak{D}_{\mm{s}}}\cap \Sigma(t) }
 g \llbracket \rho \rrbracket x_3 \nu \mm{d}S,$$
where $\nu$ is the unit inner normal on the boundary of $\mathfrak{D}_{\mm{s}}$. Thus, a partial integration results in
\begin{equation*}
G_{\mathfrak{D}_{\mm{s}}}= (0,0,-g \llbracket \rho \rrbracket |\mathfrak{D}_{\mm{s}} |)^{\mm{T}},
\end{equation*}
where $|\mathfrak{D}_{\mm{s}}|$ denotes the volume of $\mathfrak{D}_{\mm{s}}$.
The above formula reveals that the value of total pressure difference acting on the sinking part in the domain $\mathfrak{D}_{\mm{s}}$ is just equal to
the difference between the weight of sinking part of heaver fluid and that of the lighter fluid, which is displaced by the sinking part.
Moreover, the direction of total pressure difference is just along the negative direction of $x_3$-axis.
Obviously, the total pressure $G_{\mathfrak{D}_{\mm{s}}}$ further drives the heavier fluid below the interface ${\Sigma}(t)$
to sink due to the increasing volume of the sinking part of the heavier fluid.

Similarly, if we consider some part of the lower lighter fluid at time $t$ which rises above the interface ${\Sigma}(t)$, and assume that the
domain $\mathfrak{D}_{\mm{r}}$ occupied by the rising part is just a bounded connected domain,
which is bounded by the plane ${\Sigma}$ and the surface $\Sigma(t)$.
Then the total pressure difference $G_{\mathfrak{D}_{\mm{r}}}$ acting on the upper surface of
$\mathfrak{D}_{\mm{r}}$ can be represented by the following integral formula:
\begin{equation*}
G_{\mathfrak{D}_{\mm{r}}}= (0,0,g \llbracket \rho \rrbracket |\mathfrak{D}_{\mm{r}} |)^{\mm{T}}.
\end{equation*}
This means that the total pressure $G_{\mathfrak{D}_{\mm{r}}}$ further drives the lighter fluid above the interface ${\Sigma}(t)$ to rise up.
 The previous analysis is consistent with the early growth of the RT instability under the force $G_{g\llbracket\rho\rrbracket }$.

In addition, if we denote the total  pressure  difference acting on $\Sigma(t)$ by $P_{\mm{d}}(t)$, then
$P_{\mm{d}}(t)$ can be given by
$$P_{\mm{d}}(t)=\int_{\Sigma(t)}G_{g\llbracket \rho \rrbracket } \mm{d}S.$$
We evaluate that $P_{\mm{d}}(t)=0$. In fact, we have after a straightforward calculation that
$$\frac{\mm{d}}{\mm{d}t}\int_{\Omega_-(t)}\mm{d}x=\int_{\mathbb{T}^2}d_t\mm{d}x_{\mm{h}}=\int_{\mathbb{T}^2}
v\cdot (-\partial_1 d_1,-\partial_2d,1)^{\mm{T}}\mm{d}x_{\mm{h}}=\int_{\Sigma(t)}
v\cdot \nu\mm{d}x_{\mm{h}} =\int_{\Omega_-(t)} \mm{div}v\mm{d}x  =0,$$
where $\nu$ denotes the unit outer normal vector of $\partial\Omega_-(t)$. Therefore,
$$\int_{\Omega_-(t)}\mm{d}x=4\pi^2 L_1L_2 ,$$
which yields
$$P_{\mm{d}}(t)=\int_{\Sigma(t)}  g \llbracket \rho \rrbracket x_3 \nu \mm{d}S= g \llbracket \rho \rrbracket \left(0,0,
4\pi^2 L_1L_2 l- \int_{\Omega(t)}\mm{d}x \right)e_3=0.$$

Finally, we discuss the work done by the pressure difference $G_{g \llbracket \rho \rrbracket}$ acting on all particles on the interface
from time $0$ to $t$. Denote the work done by $\mathscr{W}_{g\llbracket \rho \rrbracket}(t)$.
Then, similarly to \eqref{20170712113}, we should have
\begin{equation}   \label{201711022033}
-\mathscr{W}_{g\llbracket \rho \rrbracket}(t)=\delta_{g\llbracket \rho \rrbracket}(t).
\end{equation}
We can derive the above formula by an infinitesimal method.
In fact, considering a particle $Y$ labeled by $(y_{\mm{h}},0)^{\mm{T}}$ on the interface deviating from its initial location to a new location,
then the pressure difference will do work to $Y$ in the motion process. To evaluate $\mathscr{W}_{g\llbracket \rho \rrbracket}(t)$, we denote
the location of $Y$ at time $t$ and the area element of $(y_{\mm{h}},0)^{\mm{T}}$ by $\zeta(y_{\mm{h}},0,t)$ and $\mm{d}y_{\mm{h}}$, respectively.
When the particle slightly moves to the location $\zeta(y_{\mm{h}},0,t)$, the area element of $Y$ will become $|\mathcal{A}e_3|\mm{d} y_{\mm{h}}$.
Thus, the work done by the pressure difference on $Y$ from time $\tau$ to $\tau+\mm{d}\tau$ in Lagrangian coordinates can be given by
$$ g\llbracket\rho\rrbracket\zeta_3(y_{\mm{h}},0,t) {\mathcal{A}|_{y=(y_{\mm{h}},0)}e_3}\cdot u(y_{\mm{h}},0,\tau)\mm{d}\tau\mm{d}y_{\mm{h}}. $$
Integrating the above identity over $(0,t)$, we obtain the work done by the pressure difference on $Y$ from time $0$ to $t$. Denote this work by $\mathscr{W}_{g\llbracket \rho \rrbracket}^{y_{\mm{h}}}(t)$ for the sake of simplicity.
Finally, an integration of $\mathscr{W}_{g\llbracket \rho \rrbracket}^{y_{\mm{h}}}(t)$ over $\mathbb{T}^2$ immediately yields
$\mathscr{W}_{g\llbracket \rho \rrbracket}(t)= \mathscr{V}_{g\llbracket \rho \rrbracket}^{*}(t)$.
Therefore, \eqref{201711022033} follows from \eqref{201712102027}.

\section{Proof of Theorems \ref{201707191444}--\ref{201707191444n}}\label{2017111222010}

In this section we rigorously prove Theorems \ref{201707191444}--\ref{201707191444n} stated in Section \ref{201706191309}.
We first give some preliminary results in Section \ref{201705301216}, then prove two assertions of Theorem \ref{201707191444}
in Sections \ref{201711241354}--\ref{201711241354f}. Finally we mention how to show Theorem \ref{201707191444} in Section \ref{201711241354f1907}.

\subsection{Preliminary results}\label{201705301216}


Next, we want to establish Lemma \ref{lem:modfied} below, which shows that any $\eta\in H_\sigma^1(\Omega_a^b)\cap H^{k+2}(\Omega_a^b)$ can be modified to a new function
that belongs to $H_{0,*}^{k,1}(\Omega_a^b)$ and is close to $\eta$. For this purpose, we first recall the following lemma on the global existence of inverse functions.
\begin{lem}
\label{201803121937}
Let $N\geqslant 2$, $\mathcal{D}\subseteq \mathbb{R}^N$ be an open set, $\zeta:\mathcal{D}\to \mathbb{R}^N$
belongs to $C^1(\mathcal{D})$, and $\nabla \zeta(x)$ be invertible for all $x\in \mathcal{D}$. Suppose that
$K$ is a connected compact subset of $\mathcal{D}$, and $\zeta:\partial K\to \mathbb{R}^N$ is injective, Then $\zeta :K\to \mathbb{R}^N$ is injective.
\end{lem}
We are now able to use the above lemma and the classical theory for the Stokes problem to establish the following desired lemma:
\begin{lem}
\label{lem:modfied}
Let $a<b$, $k \geqslant 1$,
If $\eta\in H^1_\sigma(\Omega_a^b)\cap H^{k+2}(\Omega_a^b)$, then there exists a constant $\varepsilon_0$ (depending on $k$ and $\eta$),
such that for any $\varepsilon\in(0,\varepsilon_0)$, there is an $\eta^{\mm{r}}\in H^1_0(\Omega_a^b)\cap H^{k+2}(\Omega_a^b)$ satisfying
$$ \det\nabla  \zeta=1 $$
and
\begin{equation}
\label{201803121036}
 \|\eta^{\mm{r}}\|_{k+2,\Omega_a^b}\leqslant c,
\end{equation}
where $\zeta:=y+\varepsilon\eta+\varepsilon^2\eta^{\mm{r}}$, the constant $c>0$ depends on $k$, the norm $\|\eta\|_{k+2,\Omega_a^b}$ and the domain $\Omega_a^b$, but not on $\varepsilon$.
Moreover,
\begin{equation}
\label{201803121601}
\zeta : \overline{\Omega_a^b\!\!\!\!\!\!\!-}\ \to \overline{\Omega_a^b\!\!\!\!\!\!\!\!-}\ \mbox{ is a homeomorphism mapping}.
\end{equation}
\end{lem}

\begin{pf} Throughout the proof, the letter $c$ denotes a general positive constant, which may depend on $k$,
the norm $\|\eta\|_{k+2,\Omega_a^b}$ and the domain $\Omega_a^b$, but independent of $\varepsilon$.

(1) We begin with the construction of $\eta^\mm{r}$. So, we consider the following Stokes problem for given $\xi\in H^3(\Omega_a^b)$:
 \begin{equation}    \label{201705141945}
 \left\{\begin{array}{ll}
-\Delta w +\nabla \varpi=0,\quad  \mm{div}w =O(\xi) &\mbox{ in }\Omega_a^b,  \\
w =0&\mbox{ on } \partial\Omega_a^b,
\end{array}\right.\end{equation}
where $O(\xi)= (1 +\varepsilon^2\mm{div}\xi-\det \nabla (y+\varepsilon\eta+\varepsilon^2\xi ))/\varepsilon^2$.
By virtue of the product estimate
$$\|fg\|_{k+1,\Omega_a^b} \leqslant c\|f\|_{k+1,\Omega_a^b} \|g\|_{k+1,\Omega_a^b}$$
with some constant $c$ depending on $k$ and $\Omega_a^b$ only, it is easy to see that
$$\|O(\xi)\|_{k+1,\Omega_a^b}\leqslant c(1+\varepsilon \|\xi\|_{k+2,\Omega_a^b}+\varepsilon^2\|\xi\|_{k+2,\Omega_a^b}^2+\varepsilon^3\|\xi\|_{k+2,\Omega_a^b}^3)\leqslant c(1 +\varepsilon^3\|\xi\|_{k+2,\Omega_a^b}^3). $$

By the classical existence and regularity theory on the Stokes problem, there exists a solution $(w,\varpi)$ of \eqref{201705141945}.
Moreover, for $\varepsilon<1$, it holds that
\begin{align}
\|w\|_{k+2,\Omega_a^b}+\|\varpi\|_{k+1,\Omega_a^b}\leqslant\|O(\xi)\|_{k+1,\Omega_a^b}\leqslant c_1 \left(1+\varepsilon^3\|\xi\|_{k+2,\Omega_a^b}^3\right),
\label{201702141356nn}  \end{align}
where the letter $c_1$ denotes a fixed constant depending on $k$, $\|\eta\|_{k+2,\Omega_a^b}$ and $\Omega_a^b$, but not on $\varepsilon$.

Therefore, one can construct an approximate function sequence $\{\eta^{n}\}_{n=1}^\infty$, such that for any $n\geqslant 2$,
\begin{equation}  \label{201702051905}
 \left\{\begin{array}{ll}
-\Delta \eta^{n}+\nabla \varpi^n=0,\quad \mm{div}\eta^{n}
=O(\eta^{n-1})&\mbox{ in }\Omega_a^b, \\
\eta^{n}=0 &\mbox{ on }\partial\Omega_a^b,  \end{array}\right.\end{equation}
where $\|\eta^{1}\|_{k+2,\Omega_a^b}\leqslant 2c_1$. Moreover, from \eqref{201702141356nn} one gets
$$ \|\eta^{n} \|_{k+2,\Omega_a^b}+ \|\varpi^n\|_{k+1,\Omega_a^b}\leqslant  c_1 (1+\varepsilon^3\|\eta^{n-1}\|_{k+2,\Omega_a^b}^3)
\quad\mbox{for any }n\geqslant 2, $$
 which implies
 \begin{equation}   \label{201702090854}
 \|\eta^{n} \|_{k+2,\Omega_a^b}+ \|\varpi^n \|_{k+1,\Omega_a^b}\leqslant 2c_1
 \end{equation} for any $n\geqslant 2$, and any $\varepsilon\leqslant 1/2c_1$.

Next we want to show that $\{\eta^n\}_{n=1}^\infty$ is a Cauchy sequence in $H^3(\Omega_a^b)$. Noting that
\begin{equation*}     \left\{\begin{array}{ll}
-\Delta( \eta^{n+1} - \eta^{n} )+\nabla (\varpi^{n+1}-\varpi^n)=0,\quad \mm{div}(\eta^{n+1}  -\eta^{n} )
=O(\eta^{n} )-O(\eta^{n-1} )&\mbox{ in }\Omega_a^b, \\
\eta^{n+1} -\eta^{n} =0 &\mbox{ on }\partial\Omega_a^b,
\end{array}\right.\end{equation*}
we obtain
\begin{equation}  \label{Ellipticestimate0837}
 \|\eta^{n+1} -\eta^{n}\|_{k+2,\Omega_a^b}+\|(\varpi^{n+1}-\varpi^n)\|_{k+1,\Omega_a^b}\leqslant c\|O(\eta^{n})-O(\eta^{n-1})\|_{k+1,\Omega_a^b}.
\end{equation}
On the other hand, using \eqref{201702090854} and the product estimates, we arrive at
 $$\|O(\eta^{n} )-O(\eta^{n-1} )\|_{k+1,\Omega_a^b}\leqslant c \varepsilon\| \eta^{n}  - \eta^{n-1}  \|_{k+2,\Omega_a^b}.$$
Substituting the above inequality into \eqref{Ellipticestimate0837}, one sees by taking $\varepsilon$ appropriately small
that $\{(\eta^n,\varpi^n)\}_{n=1}^\infty$ is a Cauchy sequence in $H^{k+2}(\Omega_a^b)\times H^{k+1}(\Omega_a^b)$.
Consequently, we can take to the limit in \eqref{201702051905} as $n\to\infty$ to see that the
limit function $(\eta^{\mm{r}},\varpi)$ solves
\begin{equation}    \label{20170514}
 \left\{\begin{array}{ll}
-\Delta \eta^{\mm{r}} +\nabla \varpi=0,\quad  \mm{div}\eta^{\mm{r}} =O(\eta^{\mm{r}}) &\mbox{ in }\Omega_a^b,  \\
\eta^{\mm{r}} =0&\mbox{ on } \partial\Omega_a^b.
\end{array}\right.  \end{equation}
 Furthermore, $\|\eta^\mm{r}\|_{k+2,\Omega_a^b}\leqslant 2c_1$ by \eqref{201702090854}.

 (2) We proceed to the proof
 \begin{equation}
\label{201804121545}
 \zeta :\overline{\Omega_a^b\!\!\!\!\!\!\!-\ }\to \overline{\Omega_a^b\!\!\!\!\!\!\!-\ }\mbox{ is injective}.
\end{equation}

Since $k\geqslant 1$ and $(\zeta-y)\in H^{1}_0(\Omega_a^b)\cap H^{k+2}(\Omega_a^b)$,
by the Sobolev embedding theorem in \cite[Section 1.3.5.8]{NASII04}, one sees that
$(\zeta-y)\in L^\infty(\Omega_a^b\!\!\!\!\!\!\!-\ ) \cap C^1(\overline{\Omega_a^b\!\!\!\!\!\!\!-\ })$ and $\zeta|_{\partial\Omega_a^b}=0$.
Let $S:=[0,2\pi L_1]\times [0,2\pi L_2]\times [a,b]$. We choose a function $\chi\in C_0^\infty(\mathbb{R}^3)$ such that
$\chi=1$ in $S$, and $0\leqslant \chi\leqslant 1$ in $\mathbb{R}^3$.
Then there exists a ball $B_{R}(0)$ of radius $R$ and center $0$, such that
$\chi=0$ in $\mathbb{R}^3 \backslash B_{R}(0)$ and $S\subset B_{R}(0)$.
Let $\zeta_\chi:=(\zeta-y)\chi/\varepsilon$, where $\zeta:=y+\varepsilon\eta+\varepsilon^2\eta^{\mm{r}}$. Then
$\zeta_\chi \in H^{k+2}(\Omega_a^b\!\!\!\!\!\!\!-\ )\cap C^1(\overline{\Omega_a^b\!\!\!\!\!\!\!-\ })$.

{Since $\Omega_a^b\!\!\!\!\!\!\!-\ $ is locally Lipschitz} (see \cite[Section 4.9]{ARAJJFF} for the definition),
by virtue of the well-known Stein extension theorem (see \cite[Section 5.24]{ARAJJFF}), there is an extension operator for
 $\Omega_a^b\!\!\!\!\!\!\!-\ $, such that
\begin{equation}
\label{20180312}
E(\zeta_\chi)=\zeta_\chi\mbox{ a.e. in }\Omega_a^b\!\!\!\!\!\!\!-\ ,
 \quad \|E(\zeta_\chi)\|_{{k+2},\mathbb{R}^3}\leqslant\tilde{c}\|\zeta_\chi\|_{k+2,\Omega_a^b\!\!\!\!\!\!\!-\ },
 \end{equation}
 where the constant $\tilde{c}$ depends on $k$ and $\Omega_a^b\!\!\!\!\!\!\!-\ $.

 Define $\zeta_{\mm{m}}:=y+ \varepsilon \chi E(\zeta_\chi)$, then
\begin{equation}   \label{201803121926}
\zeta_{\mm{m}} =\zeta\mbox{ in } S .
\end{equation}
From \eqref{201803121036}, \eqref{20180312} and the periodicity of $\eta$ and $\eta^{\mm{r}}$, it follows that
 $$\| E(\zeta_\chi)\|_{k+2,\mathbb{R}^3}\leqslant\tilde{c}\|\chi(\eta+\varepsilon \eta^{\mm{r}})\|_{k+2, \Omega_a^b\!\!\!\!\!\!\!-\ }\leqslant c,$$
 where the constant $c>0$ depends on $k$ and the norms of $\eta$ and $\eta^{\mm{r}}$, but not on $\varepsilon$. Thus, in terms of the Sobolev embedding theorem,
 $\nabla \zeta_{\mm{m}}$ is invertible in $\mathbb{R}^3$ for sufficiently small $\varepsilon$.

Let $K:=\overline{\Omega_a^b\!\!\!\!\!\!\!\!-\ \cap B_{R}(0)}$, then
$K$ is a connected compact set. Since $\zeta_{\mm{m}}(y)=y$ on $\partial K$, then $\zeta_{\mm{m}}:\partial K\to \mathbb{R}^N$
 is injective. Thus, $\zeta_{\mm{m}} :K\to \mathbb{R}^N$ is injective by Lemma \ref{201803121937}.
 Noting that $S\subset K$, we see by \eqref{201803121926} that $\zeta:S\to \mathbb{R}^N$ is also injective, which implies that $\zeta:\overline{\Omega_a^b\!\!\!\!\!\!\!-\ }\to \mathbb{R}^N$ is injective.

Now, we further show that $\zeta:\overline{\Omega_a^b\!\!\!\!\!\!\!-\ }\to \overline{\Omega_a^b\!\!\!\!\!\!\!-\ }$.
To this end, it suffices to prove that, for any $x_{\mm{h}} \in [0,2\pi L_1]\times [0,2\pi L_2]$,
$$\zeta: l_{x_{\mm{h}}}\to \overline{\Omega_a^b\!\!\!\!\!\!\!-\ },$$
where $l_{x_{\mm{h}}} :=\{(x_{\mm{h}},x_3)~|~x_3\in (a,b) \}$. We prove this by contradiction.

Assume that there exists a point $x^0\in l_{x_{\mm{h}}}$, such that $\zeta( x^0)\in  \!\!\!\!\!/ \  \overline{\Omega_a^b\!\!\!\!\!\!\!-\ }$. Without loss of generalization, we assume that $\zeta( x^0)$ is above the closed domain $\overline{\Omega_a^b\!\!\!\!\!\!\!-\ }$.
Recalling $\zeta=y$ on $\partial\Omega_a^b\!\!\!\!\!\!\!-\ $, one has $x_3^0\in (a, b)$, where $x^0_3$ denotes the third component of $x^0$.
Obviously, the curve function $\zeta$ defined on $l_0:=\{x \in l_{x_{\mm{h}}}~|~x_3\in (a,x^0_3)\}$ must go through the upper boundary of $\{x_3=b\}$.
We denote by $z$ the intersection of the curve and the upper boundary. This means that there is a point
$y\in l_0$, such that $y_3\in (a,b)$ and $\zeta(y)=z$.
Noting that $\zeta(z)=z$, we have $\zeta(y)= \zeta(z)$ where $y,z\in S$ and $y\neq z$. {This contradicts with injectivity of $\zeta$ on $S$.}
Hence \eqref{201804121545} holds.

(3) Now we turn to prove that
\begin{equation}
\label{201804121545dfadf}
\zeta :\overline{\Omega_a^b\!\!\!\!\!\!\!-\ }\to \overline{\Omega_a^b\!\!\!\!\!\!\!-\ }\mbox{ is surjective}.
\end{equation}

We define that
$$\mathbb{K}:=\{x\in \overline{\Omega_a^b\!\!\!\!\!\!\!-\ } ~|~x\neq \zeta(y)\mbox{ for any }y\in \overline{\Omega_a^b\!\!\!\!\!\!\!-}\ \}.$$
By \eqref{201804121545}, $\mathbb{K}\subset  {\Omega_a^b\!\!\!\!\!\!\!-}\ $. Obviously, to get \eqref{201804121545dfadf}, it suffices to prove that
\begin{equation}
\label{201804151053}
\mathbb{K}=\emptyset.
\end{equation}
Next we verify \eqref{201804151053} by contradiction.

We assume that $\mathbb{K}\neq \emptyset$. Noting that
$\mathbb{K}\subset \Omega_a^b\!\!\!\!\!\!\!- \ $ by \eqref{201804121545}, then $\emptyset \neq\partial \mathbb{K}\subseteq \overline{\Omega_a^b\!\!\!\!\!\!\!-}\ $.
Moreover there exists a point
\begin{equation}
\label{201804151110}
x^0\in \partial \mathbb{K}\cap \Omega_a^b\!\!\!\!\!\!\!\!-\ .
\end{equation}
Then, there exists a ball $B_{\delta_0}(x^0)\subset {\Omega_a^b\!\!\!\!\!\!\!-}\ $, such that, for any $B_{\delta}(x^0)\subset B_{\delta_0}(x^0)$,
\begin{equation}
\label{201804151502}
B_{\delta}(x^0)\cap \mathbb{K}\neq\emptyset
\end{equation}
and
\begin{equation}
\label{201804152200}
B_{\delta}(x^0)\cap ( {\Omega_a^b\!\!\!\!\!\!\!-}\ \setminus \mathbb{K})\neq\emptyset.
\end{equation}
By \eqref{201804152200}, we can choose sequences $\{x^n\}_{n=m}^\infty\subset ({\Omega_a^b\!\!\!\!\!\!\!-\ } \setminus \mathbb{K})$ and $\{y^n\}_{n=m}^\infty\subset{\Omega_a^b\!\!\!\!\!\!\!-\ }$ for some $m>1/\delta_0$, such that $x^n\in B_{1/n}(x^0)\subset B_{\delta_0}(x^0) $, $x^n\to x^0 $ and $x^n =\zeta(y^n)$. Since $(\zeta-y)\in  L^\infty(\Omega_a^b\!\!\!\!\!\!\!-\ ) \cap C^1(\overline{\Omega_a^b\!\!\!\!\!\!\!-\ })$, then
 $\{y^n\}_{n=m}^\infty$ is a bounded sequence. Therefore, there exists a subsequence (still labeled by $y^n$)
such that $y^n\to y^0$ for some $y^0\in \overline{\Omega_a^b\!\!\!\!\!\!\!-\ }$. By the continuity of $\zeta$,  $x^n=\zeta(y^n)\to x^0=\zeta(y^0)$. Since $x^0\in \Omega_a^b\!\!\!\!\!\!\!-\ $, then
$y^0\in {\Omega_a^b\!\!\!\!\!\!\!-}\ $.
Noting that $\nabla \zeta$ is invertible for $y^0$,
thus, by the well-known inverse function theorem \cite[Theorem 9.24]{WalterRudin}, 
there exist two open sets $U$ and $V\subset {\Omega_a^b\!\!\!\!\!\!\!-}\ $ such that
$y^0\in U$, $x^0\in V$, and  $\zeta(U)=V$, which imply that there exists a ball
$B_{\delta_1}(x^0)\subset  ( {\Omega_a^b\!\!\!\!\!\!\!-}\ \setminus \mathbb{K})$ for $\delta_1<\delta_0$, i.e. $B_{\delta_1}(x^0)\cap \mathbb{K}=\emptyset$, which contradicts with to \eqref{201804151502}. Hence \eqref{201804151053} holds.


(4) Finally, since $\zeta :\overline{\Omega_a^b\!\!\!\!\!\!\!-\ }\to \overline{\Omega_a^b\!\!\!\!\!\!\!-\ }$ is bijective, we can consider the inverse mapping $\zeta^{-1}$ defined on $\overline{\Omega_a^b\!\!\!\!\!\!\!-\ }$
  by \begin{equation}
  \label{201804162334}
\zeta^{-1}(\zeta(y))=y\mbox{ for }y\in \overline{\Omega_a^b\!\!\!\!\!\!\!-\ }.
 \end{equation}
 Obviously, $\zeta^{-1} :\overline{\Omega_a^b\!\!\!\!\!\!\!-\ }\to \overline{\Omega_a^b\!\!\!\!\!\!\!-\ }$ is bijective. Next we verify that
\begin{equation}
\label{201804151826}
\zeta^{-1}\mbox{ is a continuous mapping of }\overline{\Omega_a^b\!\!\!\!\!\!\!-\ }\mbox{ onto } \overline{\Omega_a^b\!\!\!\!\!\!\!-\ }
\end{equation}
by contradiction.

We assume that $\zeta^{-1}$ is not continuous for some $x^0\in \overline{\Omega_a^b\!\!\!\!\!\!\!-\ }$. Then, there exists a constant $\varepsilon>0$, such that, for any $\iota>0$, there exists a point $x^\iota\in \overline{\Omega_a^b\!\!\!\!\!\!\!-\ }$ satisfying
$|x^\iota-x^0|<\iota$ and
\begin{equation}
\label{201804151826xx}
|\zeta^{-1}(x^\iota)- \zeta^{-1}(x^0)|\geqslant  \varepsilon.
\end{equation}
Let $\iota=1/n$, we denote $y^n=\zeta^{-1}(x^{1/n})$.  Since $(\zeta-y)\in  L^\infty(\Omega_a^b\!\!\!\!\!\!\!-\ ) \cap C^1(\overline{\Omega_a^b\!\!\!\!\!\!\!-\ })$, and $\{x^{1/n}\}_{n=1}^\infty\subset B_1(x^0)\cap \overline{\Omega_a^b\!\!\!\!\!\!\!-\ }$, then $\{y^n\}_{n=1}^\infty$ is a bounded sequence. Thus there exists a subsequence (still labelled by $y^n$) such that $y^n \to y^0$ for some $y^0 \in \overline{\Omega_a^b\!\!\!\!\!\!\!-\ }$. By the continuity of $\zeta$, $x^{1/n}=\zeta(y^n)\to \zeta( y^0)=x^0$. Thus $ \zeta^{-1}(x^{1/n})\to \zeta^{-1}(x^0)$, which contracts with \eqref{201804151826xx}. Hence \eqref{201804151826} holds. Consequently, we obtain \eqref{201803121601} from \eqref{201804121545}, \eqref{201804121545dfadf} and \eqref{201804151826}. This completes
the proof of \eqref{lem:modfied}. \hfill $\Box$
\end{pf}

In addition, we can slightly modify the proof of Lemma \ref{lem:modfied} to get the following conclusion for the stratified case:
\begin{lem}
\label{lem:modfieddasfda}
Let $-l<\tau$, $k \geqslant 1$,
If $\eta\in H^1_\sigma(\Omega_{-l}^h)\cap H^{k+2}(\Omega)$, then there exists a constant $\varepsilon_0$ (depending on $k$ and $\eta$),
such that for any $\varepsilon\in(0,\varepsilon_0)$, there is an $\eta^{\mm{r}}\in H^1_0(\Omega_{-l}^h)\cap H^{k+2}(\Omega)$ satisfying
$$ \det\nabla  \zeta=1\mbox{ in }\Omega $$
and
\begin{equation*}
 \|\eta^{\mm{r}}\|_{k+2,\Omega}\leqslant c,
\end{equation*}
where $\zeta:=y+\varepsilon\eta+\varepsilon^2\eta^{\mm{r}}$, the constant $c>0$ depends on $k$, the norm $\|\eta\|_{k+2,\Omega}$ and the domain $\Omega$, but not on $\varepsilon$.
Moreover,
\begin{equation*}
\zeta : \overline{\Omega_{-l}^h\!\!\!\!\!\!\!\!\!-\ }\ \to \overline{\Omega_{-l}^h\!\!\!\!\!\!\!\!\!-\ }\ \mbox{ is a homeomorphism mapping}.
\end{equation*}
\end{lem}

Now we turn to establish an equivalence lemma of threshold, which shows that the term $\mathscr{V}_{e_j}^{\Omega}(w)$ in the definition of
$m_{\mm{N}}$ can be replaced by $\mathscr{V}_{\vec{n}}^{\Omega}(w)$, where the third component of $\vec{n}$ is $1$. To get the desired conclusion,
we shall first derive an auxiliary result.
 \begin{lem}\label{201711241502}
Let $a<b$,  and $\bar{\rho}\in C^1[a,b]$ satisfy  \begin{align}\label{0102sdafdf} \bar{\rho}'|_{s=s_0}>0\quad\mbox{for some }\; s_0\in [a,b],
\end{align}
Then there exists a function $\psi_0\in  H^1_0(a,b)$, such that $\bar{\rho}' \psi_0\neq 0$, $\psi'_0 \neq 0$, $\|\psi'_0\|_{L^2(a,b)}=1$, and
\begin{equation*}
\int_{a}^b \bar{\rho}' \psi^2_0\mm{d}s={\sup_{\psi\in H^1_0(a,b)} \frac{ \int_a^b\bar{\rho}'\psi^2\mm{d}s }{\|\psi'\|^2_{L^2(a,b)}}}=:\gamma .
\end{equation*}
\end{lem}
\begin{rem}
When $\bar{\rho}'$ is a positive constant, $\gamma$ reduces to (see \cite{JFJSJMFMOSERT})
\begin{equation}
 \label{201704070903}{\sup_{\psi\in H^1_0(a,b)} \frac{ \|\sqrt{\bar{\rho}'}\psi\|^2_{L^2(a,b)} }{\|\psi'\|^2_{L^2(a,b)}}}
  = \frac{(b-a)^2\bar{\rho}' }{ \pi^2}.      \end{equation}
\end{rem}
\begin{pf}
In view of the RT condition \eqref{0102sdafdf} and the definition of $\gamma$, we see that $\gamma>0$. Thus there is a function sequence $\{\psi_m\}_{m=1}^\infty$ satisfying $0\neq \psi_m \in H_0^1(a,b)$ and
\begin{equation*}
\frac{\int_{a}^b \bar{\rho}' \psi^2_m\mm{d}s}{\|\psi'_m\|^2_{L^2(a,h)}} \to \gamma\mbox{ as }m\to \infty .
\end{equation*}
Let $\tilde{\psi}_m:=\psi_m/\|\psi'_m\|_0$, then  $\|\tilde{\psi}'_m\|_0=1$ and $\|\tilde{\psi}_m\|_0\leqslant c$, where $c$ is independent of $m$.
Thus, there are a subsequence of $\{\tilde{\psi}_m\}_{m=1}^\infty$ (still denoted by $\tilde{\psi}_m$) and a function $\psi_0\neq 0$, such that
$$\tilde{\psi}'_m\to \psi'_0\mbox{ weakly in }H^1_0(a,b)\;\mbox{ and }\; {\int_{a}^b \bar{\rho}' \tilde{\psi}^2_m\mm{d}s} \to {\int_{a}^b \bar{\rho}' \psi^2_0\mm{d}s},$$
which gives ${\int_{a}^h \bar{\rho}' \psi^2_0\mm{d}s}=\gamma$. Hence, $\bar{\rho}'\psi_0\neq 0$ and $\psi_0'\neq 0$.
In addition, by the weakly lower semi-continuity,
$$\|\psi'_0\|_{L^2(a,b)}^2 \leqslant \liminf_{m\to \infty}\|\tilde{\psi}'_m\|_{L^2(a,b)}^2=1,$$
which, by recalling the definition of $\gamma$, implies that
$$\gamma= \int_a^b\bar{\rho}'\psi^2_0\mm{d}s  \leqslant  \frac{ \int_a^b\bar{\rho}'\psi^2_0\mm{d}s }{\|\psi'_0\|^2_{L^2(a,b)}}\leqslant \gamma. $$
Hence ${\|\psi'_0\|^2_{L^2(a,b)}}=1$. This completes the proof of Lemma \ref{201711241502}.
 \hfill $\Box$
\end{pf}
Next we further use the above result to
establish the desired conclusion.
\begin{lem}\label{addition1}
Let $a<b$,  $\bar{\rho}\in C^1[a,b]$ satisfy \eqref{0102sdafdf}, and
the third component of the constant vector $\vec{n}$ be $1$, then we have
\begin{equation}
 \label{201704070859}
b_{\mm{N}}:= {\sup_{w\in H_\sigma^{1}(\Omega_{a}^b) } \frac{\mathscr{V}_{g\bar{\rho}'}^{\Omega_a^b}(w_3)}{\mathscr{V}_{\vec{n}}^{\Omega_a^b}(w)}}=  {\sup_{w\in H_0^{1}(\Omega_{a}^b) } \frac{\mathscr{V}_{g\bar{\rho}'}^{\Omega_a^b}(w_3)}{\mathscr{V}_{\vec{n}}^{\Omega_a^b}(w)}}=  {\sup_{\psi\in H_0^1(a,b) } \frac{ g\int_{a}^b\bar{\rho}' \psi^2\mm{d}y_3 }{\lambda \|\psi'\|^2_{L^2(a,b)}}}=:c_{\mm{N}}.
 \end{equation}
\end{lem}
\begin{rem}\label{201712161034}
In view of \eqref{201704070903} and \eqref{201704070859}, we know that for any $w\in H_0^1(\Omega_+)$,
\begin{equation}    \label{201712032109}
\|\varpi\|_{0,\Omega_+}^2 \leqslant \frac{h^2}{\lambda \pi^2} \mathscr{V}_{\vec{n}}^{\Omega_+}(\varpi).
\end{equation}
\end{rem}
\begin{proof}

We prove Lemma \ref{addition1} by four steps.

(1) First of all, we claim that
\begin{equation}
 \label{201608061546ndfgdsg}
 {\sup_{w\in H_0^{1}(\Omega_{a}^b)} \frac{\mathscr{V}_{g\bar{\rho}'}^{\Omega_a^b}(w_3)}{\mathscr{V}_{\vec{n}}^{\Omega_{a}^b}(w)}}
 \leqslant {\sup_{w\in H_0^{1}(\Omega_{a}^b)} \frac{\mathscr{V}_{g\bar{\rho}'}^{\Omega_a^b}(w_3) }{\mathscr{V}_{e_3}^{\Omega_{a}^b}(w)}}=:a_{\mm{N}}.
  \end{equation}
In fact, for any given $w:=w(y_1,y_2,y_3)\in H_0^{1}(\Omega_{a}^b)$, one can verify that
$$\begin{aligned}
&\tilde{w}(y_{\mm{h}},y_3):=w(y_{\mm{h}}+ y_3\vec{n}_{\mm{h}},y_3)\in H_0^{1}(\Omega_{a}^b),\\
& \mathscr{V}_{g\bar{\rho}'}^{\Omega_a^b}(\tilde{w}_3)=\mathscr{V}_{g\bar{\rho}'}^{\Omega_a^b}(w_3),\\
&  \mathscr{V}_{e_3}^{\Omega_{a}^b}(\tilde{w}) =\| \vec{n}\cdot
\nabla w(y_{\mm{h}}+ y_3\vec{n}_{\mm{h}},y_3 )\|_{0,\Omega_{a}^b}^2=\mathscr{V}_{\vec{n}}^{\Omega_a^b}(w).
\end{aligned}  $$
Thus, \eqref{201608061546ndfgdsg} holds obviously.

(2) Then we prove that
\begin{equation}    \label{201704081148dsafasdf}
a_{\mm{N}}\leqslant c_{\mm{N}}.   \end{equation}

Let $\hat{w}_3(\xi,x_3)$ be the horizontal Fourier transform of $w_3(x)\in H_0^{1}(\Omega_{a}^b)$, i.e.,
$$ \hat{w}_3(\xi,y_3)=\int_{\mathbb{T}^2}w_3(y_h,y_3)e^{-\mm{i} y_h\cdot\xi}\mm{d}y_h, \qquad \xi=(\xi_1,\xi_2), $$
then $\widehat{\partial_3  w_3} = \partial_{3} \widehat{w}_3$.
Denote $\psi(\xi,y_3):= \psi_1(\xi,y_3) + \mm{i}\psi_2(\xi,y_3)=:\hat{w}_3(\xi,y_3)$ where $\psi_1$ and $\psi_2$ are real functions.
  By the Fubini and Parseval theorems (see \cite[Proposition 3.1.16]{grafakos2008classical}), we infer that
\begin{equation}\label{0501asdfafd}  \begin{aligned}
\mathscr{V}_{g\bar{\rho}'}^{\Omega_a^b}(w_3) = &\frac{g  }{4\pi^2 L_1L_2}\sum_{\xi\in
 L^{-1}_1\mathbb{Z} \times L^{-1}_2\mathbb{Z} } \int_a^b\bar{\rho}' (\psi_1^2(\xi,y_3)+\psi_2^2(\xi,y_3))\mm{d}y_3
\end{aligned}\end{equation}
and
\begin{equation}\label{0502asdfasf}   \begin{aligned}
\mathscr{V}_{e_3}^{\Omega_a^b}(w)\geqslant \mathscr{V}_{e_3}^{\Omega_a^b}(w_3)= \frac{\lambda}{4\pi^2 L_1L_2}\sum_{\xi\in L^{-1}_1\mathbb{Z} \times L^{-1}_2\mathbb{Z} }
\int_{a}^b (|\partial_{3}\psi_1(\xi,y_3)|^2+ |\partial_{3}\psi_2(\xi,y_3)|^2) \mm{d}y_3.
\end{aligned}\end{equation}
Keeping in mind that $\psi_i\in H_0^1(a,b)  \hbox{ if }w\in {H}_0^{1}(\Omega_{a}^b)$ by the definition of $c_{\mm{N}}$, we get
\begin{equation*} \begin{aligned}
c_{\mm{N}} \lambda \int_a^b|\psi'(\xi)|^2\mm{d}y_3\geqslant g\int_a^b \bar{\rho}'(y_3) |\psi(\xi,y_3)|^2\mm{d}y_3 .
 \end{aligned}\end{equation*}
Thus, using \eqref{0501asdfafd}--\eqref{0502asdfasf}, we deduce that for any $0\neq w\in H_0^{1}(\Omega_{a}^b)$,
\begin{equation*}\begin{aligned}
\displaystyle c_{\mm{N}}\geqslant & \frac{g  \sum_{\xi\in L^{-1}_1\mathbb{Z} \times L^{-1}_2\mathbb{Z} }
 \int_a^b\bar{\rho}' |\psi(\xi,y_3)|^2 \mm{d}y_3}{\lambda \sum_{\xi\in L^{-1}_1\mathbb{Z} \times L^{-1}_2\mathbb{Z} }
\int_a^b|\partial_{3}\psi(\xi,y_3)|^2\mm{d}y_3} = \frac{ \mathscr{V}_{g\bar{\rho}'}^{\Omega_a^b}(w_3)}{\lambda\|{\partial_3 w_3}\|^2_{0,\Omega_{a}^b} }
\geqslant \frac{\mathscr{V}_{g\bar{\rho}'}^{\Omega_a^b}(w_3) }{\mathscr{V}_{e_3}^{\Omega_{a}^b}(w)},
\end{aligned}\end{equation*}
 from which \eqref{201704081148dsafasdf} follows.

(3) Finally, we show that
\begin{equation}  \label{201704081347asdfasdf}
c_{\mm{N}}\leqslant  b_{\mm{N}}.
\end{equation}

By virtue of Lemma \ref{201711241502}, there is a function $\psi_0\in H_0^1(a,b)$, such that
\begin{equation}   \label{201711071102sdaDFSADFAF}
\frac{ g\int_a^b\bar{\rho}'|\psi_0(y_3)|^2\mm{d}y_3 }{\|\psi'_0\|_{L^2(a,b)}^2}=c_{\mm{N}}.
\end{equation}
Moreover, there is a function sequence $\{\psi_{j}\}_{j=1}^\infty\subset C^\infty_0(a,b)$ satisfying
\begin{equation}
\label{20147111162153sdfdsfs} \psi_{j}\neq 0\mbox{ and } \psi_j\to  \psi_0\mbox{ in } H_0^1(a,b)\mbox{ as }j\to \infty.
\end{equation}

We only consider the cases $\vec{n}_1\neq 0$ and $|\vec{n}_{\mm{h}}|=0$, because the other case $\vec{n}_2\neq 0$ can
be dealt with similarly.  When $\vec{n}_1\neq 0$ and $|\vec{n}_{\mm{h}}|=0$, for any given integer $i$, there always exists a real number sequence
$$l_i=\left\{
        \begin{array}{ll}
   - { \vec{n}_2L_1 i}/{\vec{n}_1 L_2}, & \hbox{ if }\vec{n}_1\neq 0 \\
         0, & \hbox{ if }|\vec{n}_{\mm{h}}|=0,
        \end{array}
      \right.
$$ such that
$$r_i:=\vec{n}_1 L^{-1}_{1} l_i +\vec{n}_2 L^{-1}_2 i =0. $$

Now we define
 \begin{equation}\label{201711241535}
  v^{i,j}(y)=(0, \psi'_{j}\cos z,   L^{-1}_2 i \psi_{j}\sin z),
\end{equation}
where $z:=  L^{-1}_{1} l_i y_1+ L^{-1}_2 i y_2$.
Then $v^{i,j}(y)\in
                 C_\sigma^\infty(\Omega_a^b) $.
Moreover, we have after a direct calculation that
\begin{align}
& \mathscr{V}^{\Omega_a^b}_{g\bar{\rho}'}(v_3^{i,j})
=gL^{-2}_2 i^2 \int_a^b\bar{\rho}'|\psi_{j}|^2\mm{d}y_3\|\sin z\|_{L^2(\mathbb{T}^2)}^2, \label{2017112215022222} \\
& \|\vec{n}\cdot\nabla   {v}^{i,j}(z) \|^2_0= \|\psi''_{j}\|^2_{L^2(a,b)}\|\cos z\|_{L^2(\mathbb{T})}^2+L^{-2}_2i^2\|\psi'_{j}\|_{L^2(a,b)}^2
\|\sin z\|_{L^2(\mathbb{T}^2)}^2. \label{201711221502}
  \end{align}

Noting that
 $$  \begin{aligned}
\|\sin z\|_{L^2(\mathbb{T}^2)}^2=2\pi^2 L_1L_2-\frac{1}{2}\int_{-\pi L_1}^{\pi L_1}\int_{-\pi L_2}^{\pi L_2} \cos 2z\mm{d}y_1\mm{d}y_2
 \end{aligned}$$
and
$$\|\cos z\|_{L^2(\mathbb{T})}^2=2\pi^2 L_1L_2+\frac{1}{2}\int_{-\pi L_1}^{\pi L_1}\int_{-\pi L_2}^{\pi L_2} \cos 2z\mm{d}y_1\mm{d}y_2 ,$$
we apply the Riemann lemma (see \cite[Theorem 16.2.1]{CJXYCHSXFX})
$$\lim_{p\to \infty}\int_a^b\psi(s)\sin ps\mm{d}s=\lim_{p\to \infty}\int_a^b\psi(s)\cos ps\mm{d}s=0\;\;\mbox{ for any }\psi\in C^0[a,b],$$
to infer that
\begin{equation}
\label{201711162154}
\|\sin z\|_{L^2(\mathbb{T}^2)}^2,\  \|\cos z\|_{L^2(\mathbb{T}^2)}^2\to 2\pi^2 L_1L_2\;\mbox{ as }i\to \infty.
\end{equation}
In view of \eqref{20147111162153sdfdsfs}, and \eqref{2017112215022222}--\eqref{201711162154}, one has
\begin{align}
\label{201711210917}
b_{\mm{N}} \geqslant \lim_{j\to \infty} \lim_{i\to \infty}\frac{\mathscr{V}
_{g\bar{\rho}'}^{\Omega_a^b}(v_3^{i,j}) }{ \mathscr{V}_{\vec{n}}^{\Omega_a^b}( {v}^{i,j})}= \lim_{j\to \infty}  \frac{g\int_a^b\bar{\rho}' |\psi_j(y_3)|^2\mm{d}y_3   }{ \|\psi'_j\|_{L^2(a,b)}^2 }=c_{\mm{N}},
\end{align}
which gives \eqref{201704081347asdfasdf}.

(4) We are now in a position to show \eqref{201704070859}. From \eqref{201608061546ndfgdsg}, \eqref{201704081148dsafasdf}
and \eqref{201704081347asdfasdf}, we get
$$ {\sup_{w\in H_0^{1}(\Omega_{a}^b)} \frac{\mathscr{V}
_{g\bar{\rho}'}^{\Omega_a^b}(w_3)}{\mathscr{V}_{\vec{n}}^{\Omega_a^b}(w)}} \leqslant a_{\mm{N}}\leqslant c_{\mm{N}} \leqslant b_{\mm{N}}. $$
On the other hand, since $H_\sigma^{1}(\Omega_{a}^b)\subset H_0^{1}(\Omega_{a}^b)$, we see that
$$b_{\mm{N}} \leqslant  {\sup_{w\in H_0^{1}(\Omega_{a}^b)} \frac{\mathscr{V}
_{g\bar{\rho}'}^{\Omega_a^b}(w_3)}{\mathscr{V}_{\vec{n}}^{\Omega_{a}^b}(w)}}. $$
Consequently, we obtain the desired conclusion \eqref{201704070859}.
\end{proof}

Similarly to Lemma \ref{addition1}, we also an equivalence lemma of threshold for the stratified case.
\begin{lem}\label{addition1sdaf}
Let $h$, $l$, $g \llbracket \rho \rrbracket >0$, and  the third component of the constant vector $\vec{n}$ be $1$, then we have
\begin{equation*}
  \sup_{w\in H_\sigma^{1}(\Omega_{-l}^h)} \frac{\mathscr{V}_{g\llbracket \rho \rrbracket}(w_3)}{\mathscr{V}_{\vec{n}}^{\Omega_{-l}^h}(w)}=  {\sup_{w\in H_0^{1}(\Omega_{-l}^h)} \frac{\mathscr{V}_{g\llbracket \rho \rrbracket}(w_3) }{\mathscr{V}_{e_3}^{\Omega_{-l}^h}(w)}}=  {\sup_{\psi\in H_0^1(-l,h)} \frac{ g\llbracket \rho \rrbracket |\psi(0)|^2 }{\lambda \|\psi'\|^2_{L^2(-l,h)}}} .
 \end{equation*}
\end{lem}
\begin{rem}\label{201712121004}
  In \cite{JFJSWWWOA} Wang gave the following formula
$$\sup_{\psi\in  {H}_0^1(-l,h) } \frac{  |\psi(0)|^2 }{\|\psi'\|^2_{L^2(-l,h)}}=  (h^{-1}+l^{-1})^{-1}, $$
where the supremum can be achieved by choosing
\begin{equation*}
\psi(s)=\left\{
                              \begin{array}{ll}
1+s/l, & s\in (-l,0];\\
1-s/h, & s\in (0,h).
\end{array}  \right.
\end{equation*}
\end{rem}
\begin{pf}
We omit the trivial derivation here, since Lemma \ref{addition1sdaf} can be easily shown by slightly modifying the arguments
in the proof of Lemma \ref{addition1}.   \hfill$\Box$
\end{pf}

\subsection{Proof of the first assertion in Theorem \ref{201707191444}}\label{201711241354}

We now prove the first assertion in Theorem \ref{201707191444}. Without loss of generality, we assume that $\bar{M}_1\neq 0$.
By Lemmas \ref{201711241502} and \ref{addition1}, there exists a $\psi_0\in H_0^1(0,h)$ satisfying
   \begin{equation*}
m_{\mm{N}}=\sqrt{\frac{g\int_0^h \bar{\rho}' \psi^2_0\mm{d}s}{\lambda \|\psi'_0\|^2_{L^2(0,h)}}}.
  \end{equation*}
Since $\psi_0\in H_0^1(0,h)$ and $m_{\mm{N}}>0$, there is a function sequence $\{\psi_{j}\}_{j=1}^\infty\subset C_0^\infty(0,h)$ satisfying
\begin{equation}\label{201712140909}
\bar{\rho}'\psi_{j}\neq 0,\ \psi_{j}\neq 0
\mbox{ and }
 \psi_j\to  \psi_0\mbox{ in }H_0^1(0,h)\mbox{ as }j\to \infty.
\end{equation}

Let $i$ be a positive integer to be determined later. Let $v^{i,j}$ be defined by \eqref{201711241535} with $\psi_j$ constructed as above.
Then $v^{i,j}\in C_\sigma^\infty(\Omega_+)$. Moreover, we have after a straightforward computation that
\begin{align}
&  \mathscr{V}_{ {g\bar{\rho}'}}^{\Omega_+}(  v_3^{i,j})
= gL^{-2}_2   i^2   \int_{0}^h\bar{\rho}'|\psi_j|^2\mm{d}s\|\sin z\|_{L^2(\mathbb{T}^2)}^2\neq 0 , \label{201803152146} \\
& \mathscr{V}_{\bar{M}}^{\Omega_+}(  v^{i,j}) =  \bar{M}_3^2 \left(\|\psi''_{j}\|^2_{L^2(0,h)}\|\cos z\|_{L^2(\mathbb{T}^2)}^2+ L^{-2}_2i^2\|\psi'_{j}\|_{L^2(0,h)}^2\|\sin z\|_{L^2(\mathbb{T}^2)}^2\right) .
\label{201712112210}
\end{align}
 In what follows, the letters $c(i,j)$, $c_k(i,j)$ and $\varepsilon_1(i,j)$ denote
positive constants for $k=2$ and $3$, which may depend on $v^{i,j}$, but not on $\varepsilon\in (0,1)$.
Moreover, $c(i,j)$ can vary from line to line.

Since $v^{i,j}$ may not belong to $H^{k,1}_{0,*}(\Omega_+)$, we have to further modify it.
For $v^{i,j}\in C_\sigma^\infty(\Omega_+)$, by virtue of Lemma \ref{lem:modfied}, there is an $\varepsilon_1(i,j)\in (0,1)$,
such that for any $\varepsilon\in (0,\varepsilon_1(i,j))$, there exists an
${\varpi}^{i,j}_{\varepsilon,\mm{r}}\in H_{0}^{k}(\Omega_+) $ satisfying
\begin{equation*}
\begin{aligned}
&\|{\varpi}^{i,j}_{\varepsilon,\mm{r}}\|_{k,\Omega_+}\leqslant c(i,j),\\
& \varpi^{i,j}_\varepsilon:=\varepsilon v^{i,j}+ \varepsilon^2\varpi_{\varepsilon,\mm{r}}^{i,j}\in H_{0,*}^{k,1}(\Omega_+),
\end{aligned}
 \end{equation*}
where the constant $\varepsilon_1(i,j)>0$ may depend on $v^{i,j}$, but not on $\varepsilon$.
It is easy to see that
\begin{equation}    \label{201712112243}
 \|(v^{i,j},\varpi^{i,j}_{\varepsilon,\mm{r}} )\|_{C^1(\overline{\Omega_+})}    \leqslant  c(i,j).
\end{equation}

We further define
$$
\begin{aligned}
\mathfrak{R}_\varepsilon^{i,j}:= &  \varepsilon^3\left( g \int_{\Omega_+}\bar{\rho}' v_3^{i,j} ({\varpi}^{i,j}_{\varepsilon,\mm{r}})_3  \dx- \lambda\int_{\Omega_+}\partial_{\vec{n}} v^{i,j}\cdot\partial_{\vec{n}}  {\varpi}^{i,j}_{\varepsilon,\mm{r}}\mm{d}y\right)\\
&+\frac{\varepsilon^4}{2}(\mathscr{V}_{ {g\bar{\rho}'}}^{\Omega_+}(({\varpi}^{i,j}_{\varepsilon,\mm{r}})_3)-\mathscr{V}_{\vec{n} }^{\Omega_+}( {\varpi}^{i,j}_{\varepsilon,\mm{r}}  ))+
\mathfrak{N}_{g\bar{\rho}'}(\varpi^{i,j}_\varepsilon),
\end{aligned}$$
where $({\varpi}^{i,j}_{\varepsilon,\mm{r}})_3$ denotes the third component of ${\varpi}^{i,j}_{\varepsilon,\mm{r}}$,
$$\bar{M}^*:=\left\{
    \begin{array}{ll}
      \bar{M}/\bar{M}_3, & \hbox{ if }\bar{M}_3\neq 0, \\
     \bar{M}, & \hbox{ if }\bar{M}_3=0
    \end{array}
  \right.\mbox{ and } {\vec{n}} :=\left\{
    \begin{array}{ll}
      \bar{M}, & \hbox{ if }\bar{M}_3\neq 0, \\
     \bar{M}^*, & \hbox{ if }\bar{M}_3=0.
    \end{array}
  \right.
$$
Utilizing \eqref{201709062212} and \eqref{201712112243}, we infer that
\begin{equation}    \label{201712112247}
\left\|\mathfrak{R}_{\varepsilon}^{i,j} \right\|_{L^\infty(\Omega_+)}\leqslant c_2(i,j) \varepsilon^3.
\end{equation}

Now we consider the case $\bar{M}_3=0$, and choose a positive integer $i=i_0$ and a function $\psi_{j_0}$. Thanks to \eqref{201803152146}, one has
\begin{equation*}
 \mathscr{V}_{ {g\bar{\rho}'}}^{\Omega_+ }( \varepsilon v_3^{i_0,j_0}) \geqslant  2 c_3(i_0,j_0) \varepsilon^2,
\end{equation*}
where $c_3$ is independent of $\varepsilon$. Noting $\mathscr{V}_{\bar{M}}^{\Omega_+}(v^{i,j})=0$, we obtain
$$  \begin{aligned}
\frac{1}{2}\mathscr{V}_{\bar{M}}^{\Omega_+}({\varpi}^{i_0,j_0}_\varepsilon )+ \delta_{g\bar{\rho}'}(\varpi^{i_0,j_0}_\varepsilon)
=&  - \frac{1}{2} \mathscr{V}_{ {g\bar{\rho}'}}^{\Omega_+}(\varepsilon v_3^{i_0,j_0}) -\mathfrak{R}_\varepsilon^{i_0,j_0}\\
\leqslant &  \varepsilon^2(c_2(i_0,j_0)\varepsilon -c_3(i_0,j_0))<-\varepsilon^2c_3(i_0,j_0)/2
\end{aligned}$$
for any $\varepsilon<\varepsilon_0:=\min\{\varepsilon_1(i_0,j_0),c_3(i_0,j_0)/2c_2(i_0,j_0)\}$,
which yields $\delta_{E_{\mm{P}}}^N(\varpi^{i_0,j_0}_\varepsilon)<0$. Therefore, the first assertion in Theorem \ref{201707191444} holds
for the case $\bar{M}_3=0$.

For the case $\bar{M}_3\neq 0$, we make use of \eqref{201711210917} and \eqref{201712140909} to find that
\begin{equation}
\label{201711301018}
m_{\mm{N}}^2\geqslant \lim_{j\to \infty} \lim_{i\to \infty}\frac{\mathscr{V}_{g\bar{\rho}'}^{\Omega_+}(v_3^{i,j}) }{ \mathscr{V}_{\bar{M}^*}^{\Omega_+}( {v}^{i,j})}= \lim_{j\to \infty}  \frac{   g\int_0^h \bar{\rho}' \psi^2_j\mm{d}s    }{ \lambda\|\psi'_j\|_{L^2(0,h)}^2 }=m_{\mm{N}}^2.
\end{equation}
Then, for any $\delta>0$, there are $i_\delta$ and $j_\delta$ (depending on $\delta$), such that
$$0\leqslant m_{\mm{N}}^2- \frac{\mathscr{V}_{g\bar{\rho}'}^{\Omega_+}(v_3^{i_\delta,j_\delta}) }{ \mathscr{V}_{{\bar{M}^*}}^{\Omega_+}
( v^{i_\delta,j_\delta})} < \delta. $$

Now, let $\delta:=(m_{\mm{N}}^2-\bar{M}_3^2)/2$. Due to $\bar{M}_3<m_{\mm{N}}$, one sees
\begin{align}
\mathscr{V}_{ {g\bar{\rho}'}}^{\Omega_+}(v_3^{i_\delta,j_\delta}) > (\bar{M}_3^2+( m_{\mm{N}}^2 - \bar{M}_3^2)/2)\mathscr{V}_{{\bar{M}^*}}^{\Omega_+}(v^{i_\delta,j_\delta}).
\label{20170719new}
\end{align}
From \eqref{201712112210} one gets
 $$0< c(i_\delta,j_\delta)\varepsilon^2 \leqslant  (m_{\mm{N}}^2-\bar{M}_3^2)\mathscr{V}_{{\bar{M}^*}}^{\Omega_+}(\varepsilon v^{i_\delta,j_\delta} ).$$
Hence, by virtue of \eqref{201712112247}, there is a sufficiently small constant $\varepsilon_0(i_\delta,j_\delta)<{\varepsilon}_1(i_\delta,j_\delta)$,
such that for any $\varepsilon\in (0,\varepsilon_0(i_\delta,j_\delta))$,
\begin{align}
 \frac{m_{\mm{N}}^2-\bar{M}_3^2}{4}\mathscr{V}_{\bar{M}^*}^{\Omega_+}(\varepsilon v^{i_\delta,j_\delta} )
 +\mathfrak{R}_{\varepsilon}^{i_\delta,j_\delta}> 0.\label{20170719new201712031534}
\end{align}

If we make use of \eqref{20170719new} and \eqref{20170719new201712031534}, we derive that
$$  \begin{aligned}
-\delta_{g\bar{\rho}'}(\varpi^{i_\delta,j_\delta}_\varepsilon)
=&\frac{\varepsilon^2}{2}\mathscr{V}_{ {g\bar{\rho}'}}^{\Omega_+}(v^{i_\delta,j_\delta}_3 )
+  g \varepsilon^3 \int_{\Omega_+}\bar{\rho}' v^{i_\delta,j_\delta}_3  ({\varpi}^{i_\delta,j_\delta}_{\varepsilon,\mm{r}})_3  \dx+
\frac{\varepsilon^4}{2}\mathscr{V}_{ {g\bar{\rho}'}}^{\Omega_+}( v^{i_\delta,j_\delta}_{\varepsilon,\mm{r}} )+
\mathfrak{N}_{g\bar{\rho}'}({\varpi}^{i_\delta,j_\delta}_\varepsilon)\nonumber \\
>&\frac{\varepsilon^2}{2}(\bar{M}_3^2+( m_{\mm{N}}^2 - \bar{M}_3^2)/2)\mathscr{V}_{\bar{M}^*}^{\Omega_+}(v^{i_\delta,j_\delta})\nonumber \\
&+  g \varepsilon^3 \int_{\Omega_+}\bar{\rho}' v^{i_\delta,j_\delta}_3 (\varpi^{i_\delta,j_\delta}_{\varepsilon,\mm{r}})_3  \dx+\frac{\varepsilon^4}{2}\mathscr{V}_{ {g\bar{\rho}'}}^{\Omega_+}((\varpi^{i_\delta,j_\delta}_{\varepsilon,\mm{r}})_3)+
\mathfrak{N}_{g\bar{\rho}'}({\varpi}^{i_\delta,j_\delta}_\varepsilon)\nonumber \\
=& \frac{\bar{M}_3^2}{2}\mathscr{V}_{{\bar{M}^*}}^{\Omega_+}(\varpi^{i_\delta,j_\delta}_\varepsilon)  +\frac{ m_{\mm{N}}^2 - \bar{M}_3^2}{4}\mathscr{V}_{{\bar{M}^*}}^{\Omega_+}(\varepsilon v^{i_\delta,j_\delta})  +\mathfrak{R}^{i_\delta,j_\delta}_{\varepsilon}> \frac{1}{2}\mathscr{V}_{\bar{M}}^{\Omega_+}({\varpi}^{i_\delta,j_\delta}_\varepsilon ),
\end{aligned}
$$
which yields  $\delta_{E_{\mm{P}}}^N(\varpi^{i_\delta,j_\delta}_\varepsilon)<0$. Therefore, the first assertion in Theorem \ref{201707191444}
also holds for $\bar{M}_3\neq 0$. This completes the proof.

\subsection{Proof of the second assertion in Theorem \ref{201707191444}}\label{201711241354f}

Since $\bar{M}_3\neq 0$, we denote $\vec{n}:=\bar{M}/\bar{M}_3$.
By \eqref{201709062212}, \eqref{201709041500sdaf} and \eqref{201712032109}, we see that there is a sufficiently small constant $\varepsilon_0$,
such that for any non-zero function $\varpi\in H^1_0(\Omega^+)$ satisfying $\|\varpi\|_{L^\infty(\Omega_+)}\leqslant \varepsilon_0$, it holds that
\begin{equation*}
\mathfrak{N}_{g\bar{\rho}'}(\varpi) \leqslant  c\|\varpi\|_{L^3(\Omega_+)}^3 \leqslant c \|\varpi\|_{L^\infty(\Omega_+)}
 \mathscr{V}_{\vec{n}}^{\Omega_+}(\varpi) < \frac{\bar{M}_3^2-m_{\mm{N}}^2}{2}\mathscr{V}_{\vec{n}}^{\Omega_+}(\varpi).
\end{equation*}
On the other hand, from the definition $m_{\mm{N}}$ one gets
$$ m_{\mm{N}}^2 \mathscr{V}_{\vec{n}}^{\Omega_+}(\varpi) \geqslant \mathscr{V}_{{g\bar{\rho}'}}(\varpi). $$
Thus, for any non-zero function $\varpi\in  H^1_0(\Omega_+)$,
$$\begin{aligned}
\delta_{g\bar{\rho}'}(\varpi) \leqslant &\frac{m_{\mm{N}}^2}{2}  \mathscr{V}_{\vec{n}}^{\Omega_+}  (\varpi) +\mathfrak{N}_{ g\bar{\rho}' }(\varpi)\nonumber \\
<& \frac{m_{\mm{N}}^2}{2}  \mathscr{V}_{\vec{n}}^{\Omega_+}(\varpi) +\frac{\bar{M}_3^2-m_{\mm{N}}^2}{2}\mathscr{V}_{\vec{n}}^{\Omega_+}(\varpi) =
\frac{1}{2}\mathscr{V}_{\bar{M}}^{\Omega_+}(\varpi).
\end{aligned}
$$
Hence, the second assertion in Theorem \ref{201707191444} holds.

\subsection{Proof of Theorem \ref{201707191444n}}\label{201711241354f1907}

With the help of \eqref{201711262224567894}, Lemmas \ref{lem:modfieddasfda} and \ref{addition1sdaf}, and Remark \ref{201712121004}, we can easily establish
Theorem \ref{201707191444n} by following the arguments in the proof of Theorem \ref{201707191444}, and we omit the details here.

\section{Extension to the magnetic B\'enard problem}\label{Sec:0302}

In this section we further extend the results on the magnetic inhibition in the NMRT problem to the magnetic B\'enard problem without heat
conduction. We begin with a brief introduction of the thermal instability. Thermal instability often arises when a fluid is heated from below.
The classic example of this is a horizontal layer of fluid with its lower side hotter than its upper. The basic state is then one of rest states
with light and hot fluid below heavy and cool fluid. When the temperature difference across the layer is great enough, the stabilizing effects
of viscosity and thermal conductivity are overcome by the destabilizing (thermal) buoyancy, and an overturning instability ensues
 as thermal convection: hotter part of fluid is lighter and tends to rise as colder part tends to sink according
to the action of the gravity force \cite{DPGRWHHC}.

 The effect of an impressed magnetic field on the onset of thermal instability in MHD fluids is first considered by Thompson \cite{TWBTCIA} in 1951.
 Then Chandrasekhar theoretically further discovered the inhibiting effect of the magnetic field on the thermal instability in 1952 \cite{CSTPRSOTICB,CSTPRSOTICBII},
Later, the nonlinear magnetic inhibition theory with resistivity was given by Galdi \cite{GGPNA}, also see \cite{GGPPMFR,GGPPMNA,MGRSNSC}.
  However, by now it is still an open problem to show the inhibition of thermal instability by magnetic fields based on the following 3D magnetic
  Boussinesq equations without resistivity in $\Omega_+$ (see \cite[Chapter IV]{CSHHSCPO} for a derivation):
\begin{equation}\label{0101ooo}\left\{{\begin{array}{ll}
  \beta v_t+ \beta v\cdot \nabla v+\nabla \left(p+ \lambda|M|^2/2\right) =g \beta(\alpha(\Theta-\Theta_2)-1)e_3
  + \mu\Delta v+\lambda {M}\cdot \nabla M , \\
  \Theta_t +v\cdot\nabla \Theta =\kappa \Theta , \\[1mm]
M_t+v\cdot \nabla M=M\cdot \nabla v,\\
\mm{div}v= \mm{div }M=0 . \end{array}}  \right.
\end{equation}
We shall complementally introduce the new mathematical notations appearing in the equations \eqref{0101ooo}.
The unknown function $\Theta:=\Theta(x,t)$ represents the temperature of the incompressible MHD fluid.
The new known physical parameters $\beta$, $\alpha$ and $\kappa$ denote the density constant at some properly chosen temperature
parameter $\Theta_2$, the coefficient of volume expansion and the coefficient of heat conductivity, respectively.

However, if we omit the thermometric conductivity (i.e. $\kappa$=0), we can easily get the inhibition of thermal instability by a magnetic field.
In fact, when $\kappa$=0, the system \eqref{0101ooo} reduces to
\begin{equation}\label{0101oooafas}\left\{{\begin{array}{ll}
   \beta v_t+ \beta v\cdot \nabla v+\nabla \mathcal{P}=g \alpha \beta \Theta e_3+ \mu \Delta v+\lambda {M}\cdot \nabla M  ,\\
  \Theta_t +v\cdot\nabla \Theta =0,\\[1mm]
M_t+v\cdot \nabla M=M\cdot \nabla v,\\
\mm{div}v= \mm{div }M=0,\end{array}}  \right.
\end{equation}
where $\mathcal{P}:=p+ \lambda|M|^2/2  +g\beta (\alpha \Theta_2+1)x_3.$

For the well-posedness of \eqref{0101oooafas}, we need that following initial and boundary conditions:
\begin{align}
& \label{0104dsafafds} (v,\theta,M)|_{t=0}=( v^0,\theta^0, N^0) \quad \mbox{in } D,\\[1mm]
&  \label{0105n} v(x,t)=0 \mbox{ on }\partial D.
\end{align}
For simplicity, we call the model \eqref{0101oooafas}--\eqref{0105n} the magnetic B\'enard model without heat conduction.

The rest state of the above {magnetic B\'enard } model can be given by ${r_\mm{B}}:=(0,\bar{\Theta},\bar{M})$ with an associated equilibrium
pressure $\bar{p}$, where $\bar{\Theta}$ and $\bar{p}$ are smooth functions defined on $\overline{D}$, depend on $x_3$ only, and satisfy
the equilibrium relation
$$ \nabla \bar{p} =g \beta (\alpha(\bar{\Theta}-\Theta_2)-1)e_3$$
and the convection condition
\begin{equation}    \label{201705021821nn}
\bar{\Theta}'|_{x_3=x_3^0}<0\;\;\mbox{ for some }x^0_{3}\in {D_{x_3}}:= \{x_3~|~(x_{\mm{h}},x_3)\in D\}.  \end{equation}
If $D$ takes $\Omega^+$, we shall further assume that the density profile $\bar{\Theta}$ is a vertically horizontal function, i.e.,
$$\bar{\Theta}|_{x_3=2\pi n L_2}=\bar{\Theta}|_{x_3=2\pi m L_2}\;\;\mbox{ for any integer }n,m.$$

The convection condition \eqref{201705021821nn} assures that there is at least a region in which the temperature profile $\bar{\Theta}$
has lower temperature with increasing height $x_3$, and thus may lead to the classical convective instability \cite{CSHHSCPO}.
The problem whether ${r_{\mm{B}}}$ is stable or unstable to the {magnetic B\'enard } model without heat conduction is called
the magnetic B\'enard problem without heat conduction. We easily see that the {magnetic B\'enard} problems is extremely similar to the {NMRT} problem.
This is not surprising from the physical point of view: the thermal instability is caused by the interchange of the lower hotter fluid and the upper
cooler fluid driven by buoyancy, like the RT instability induced by the interchange of the upper heavier fluid and the lower lighter
fluid driven by gravity. Mathematically, both instabilities correspond to two terms $g\alpha \Theta e_3$ and $- g \rho e_3$ respectively.
Hence, the thermal and RT instabilities belong to the type of interchange instability.

In view of this similarity, we easily observe that all mathematically results of stability/instability for the NMRT problem
can be directly extended to the magnetic B\'enard  problem without heat conduction. In fact, if we define
$$
\mathscr{V}_{g\alpha \beta \bar{\Theta}'}^D(w_3):= {-\int_{D} g\alpha \beta \bar{\Theta}'|w_3|^2\mm{d}x} \;\;\mbox{ and }\;\;
 m_{\mm{B},j}^{D}:=\sqrt{\sup_{w\in H_{\sigma}^1(D)}\frac{\mathscr{V}_{g\alpha \beta \bar{\Theta}'}^D(w)}{\mathscr{V}_{e_j}^{D}(w)}}, $$
then, similarly to the NMRT problem, we have the following stability/instability results on the {magnetic B\'enard} problem without heat conduction.
\begin{enumerate}[\quad (1)]
\item Stability criterion: if $|\bar{M}_j|>{m}_{\mm{B},j}^{\Omega_j}$, then the rest state $r_{\mm{B}}$ with $\bar{M}=\Pi_j$ is asymptotically
stable to the {magnetic B\'enard } model defined on $D=\Omega_j$ with proper initial condition under small perturbation for $j=1$ and $3$.
\item Instability criterion: if $|\bar{M}_j|<{m}_{\mm{B},j}^{\Omega_j}$, then the rest state ${r_{\mm{B}}}$ with $\bar{M}=\Pi_j$ is unstable
to the {magnetic B\'enard} model defined on $D=\Omega_j$ in the Hadamard sense for $j=1$ and $3$.
In addition, the rest state ${r_{\mm{B}}}$ with $\bar{M}=\Pi_k$ is always unstable to the {magnetic B\'enard} model defined on $D=\Omega_j$
in the Hadamard sense, when $k\neq j$.
\end{enumerate}

The above stability result shows that the thermal instability can be inhibited by (impressed) magnetic fields in a non-resistive MHD fluid
with heat conduction. We mention that a sufficiently large $\mu$ in the system \eqref{0101ooo} can inhibit the thermal instability.
However, $\mu$ in the system \eqref{0101oooafas} can never inhibit the thermal instability due to the absence of $\kappa$. Of course,
we can rigourously prove that $\mu$ can slow down the development of the linear thermal instability in \eqref{0101oooafas} in some sense,
see \cite{JJWWW2014nF} on the effect of viscosity on the linear RT instability.

\vspace{4mm} \noindent\textbf{Acknowledgements.}
 The research of Fei Jiang was supported by NSFC (Grant No. 11671086), the NSF of Fujian Province of China (Grant No. 2016J06001)
 and the Education Department of Fujian Province (Grant No. SX2015-02), and the research of Song Jiang by the Basic Research Program (2014CB745002)
and NSFC (Grant Nos. 11631008, 11371065, 11571046). The authors thank Prof. C.H. Arthur Cheng for pointing out Lemma \ref{201803121937}.

\renewcommand\refname{References}
\renewenvironment{thebibliography}[1]{%
\section*{\refname}
\list{{\arabic{enumi}}}{\def\makelabel##1{\hss{##1}}\topsep=0mm
\parsep=0mm
\partopsep=0mm\itemsep=0mm
\labelsep=1ex\itemindent=0mm
\settowidth\labelwidth{\small[#1]}%
\leftmargin\labelwidth \advance\leftmargin\labelsep
\advance\leftmargin -\itemindent
\usecounter{enumi}}\small
\def\newblock{\ }
\sloppy\clubpenalty4000\widowpenalty4000
\sfcode`\.=1000\relax}{\endlist}
\bibliographystyle{model1b-num-names}

\end{CJK*}
\end{document}